%% file: main.tex
\documentclass{vldb}
\usepackage{balance}  
\usepackage{url}

\usepackage[T1]{fontenc}
\usepackage{tgtermes}

\usepackage[normalem]{ulem}

\usepackage[linesnumbered,ruled,vlined]{algorithm2e}
\usepackage{graphicx}
\usepackage{caption}
\usepackage{subcaption}

\usepackage{booktabs}
\usepackage{multirow}

\numberofauthors{3}
\author{
\alignauthor Yuliang Li\\
\affaddr{UC San Diego}\\
\email{yul206@eng.ucsd.edu }
\alignauthor Alin Deutsch\\
\affaddr{UC San Diego}\\
\email{deutsch@cs.ucsd.edu}
\alignauthor Victor Vianu \\
\affaddr{UC San Diego \& INRIA Saclay} \\ 
\email{vianu@cs.ucsd.edu}
}

\title{VERIFAS: A Practical Verifier for Artifact Systems}
\date{}

\newtheorem{theorem}{Theorem}

\newtheorem{definition}[theorem]{Definition}

\newtheorem{example}[theorem]{Example}
\newtheorem{remark}[theorem]{Remark}

\newtheorem{problem}[theorem]{Problem}

\usepackage{graphicx}

\usepackage[usenames]{color}
\usepackage{enumitem}
\setlist[itemize]{leftmargin=*}
\setlist[enumerate]{leftmargin=*}

\newcommand{\type}{\ensuremath{\mathtt{type}}}

\newcommand{\alin}[1]{{\it\small\textcolor{green}{[[[ {#1}\ --alin ]]]}}}
\newcommand{\victor}[1]{{\it\small\textcolor{red}{[[[ {#1}\ --victor ]]]}}}
\newcommand{\yuliang}[1]{{\it\small\textcolor{blue}{[[[ {#1}\ --yuliang ]]]}}}
\newcommand{\reviewer}[1]{{\it\small\textcolor{cyan}{[[[ {#1}\ --reviewers ]]]}}}

\renewcommand{\alin}[1]{}
\renewcommand{\victor}[1]{}
\renewcommand{\yuliang}[1]{}
\renewcommand{\reviewer}[1]{}

\newcommand{\eat}[1]{}

\newcommand{\calc}{{\cal C}}

\newcommand{\cala}{{\cal A}}
\newcommand{\calh}{{\cal H}}

\newcommand{\cals}{{\cal S}}
\newcommand{\calt}{{\cal T}}
\newcommand{\cale}{{\cal E}}
\newcommand{\cali}{{\cal I}}

\newcommand{\lora}{\longrightarrow}
\newcommand{\goto}[1]{\stackrel{#1}{\lora}}
\newcommand{\db}{\mathcal{DB}}

\newcommand{\aactive}{\ensuremath{\mathtt{active}}}
\newcommand{\aclosed}{\ensuremath{\mathtt{inactive}}}

\newcommand{\anull}{\ensuremath{\mathtt{null}}}

\newcommand{\buchi}{B\"{u}chi}

\newcommand{\varid}{\ensuremath{\emph{VAR}_{id}}}
\newcommand{\varnum}{\ensuremath{\emph{VAR}_{\emph{val}}}}

\newcommand{\dbcustomers}{\ensuremath{\mathtt{CUSTOMERS}}}
\newcommand{\dbitems}{\ensuremath{\mathtt{ITEMS}}}
\newcommand{\dbrecords}{\ensuremath{\mathtt{CREDIT\_RECORD}}}

\newcommand{\arecord}{\ensuremath{\mathtt{record}}}
\newcommand{\acustomerid}{\ensuremath{\mathtt{cust\_id}}}
\newcommand{\aitemid}{\ensuremath{\mathtt{item\_id}}}
\newcommand{\astatus}{\ensuremath{\mathtt{status}}}

\newcommand{\ainstock}{\ensuremath{\mathtt{instock}}}

\begin{document}
\maketitle

\begin{abstract}
Data-driven workflows, of which IBM's Business Artifacts are a prime exponent, have 
been successfully deployed in practice, adopted in industrial standards, and
have spawned a rich body of research in academia, focused primarily on static analysis.
The present research bridges the gap between the theory and practice 
of artifact verification with VERIFAS, 
the first implementation of practical significance of an artifact verifier 
with full support for unbounded data. VERIFAS verifies within seconds
linear-time temporal properties over real-world and synthetic workflows of complexity 
in the range recommended by software engineering practice. 
Compared to our previous implementation based on the widely-used Spin model checker,
VERIFAS not only supports a model with richer data manipulations but also outperforms it by over an order of magnitude.
VERIFAS' good performance is due to a novel symbolic representation approach and 
a family of specialized optimizations.
\end{abstract}

\keywords{data-centric workflows; business process management; temporal logic; verification}

\input{introduction}

\input{newmodel}

\input{karp-miller}
\input{experiment}

\input{related-work}

\input{conclusion}
\bibliographystyle{abbrv}
{
\bibliography{reference,artifacts,art2}
}

\appendix

\input{appendix-semantics}

\input{appendix-example}

\input{appendix-karp-miller}
\input{appendix-experiment}

\input{appendix-table}

\end{document}

%% file: introduction.tex
\section{Introduction} \label{sec:intro}
The past decade has witnessed the evolution of workflow specification frameworks from the traditional
process-centric approach towards data-awareness. Process-centric formalisms focus on control flow 
while under-specifying the underlying data and its manipulations by the process
tasks, often abstracting them away completely. In contrast, data-aware formalisms treat data as first-class
citizens. A notable exponent of this class is IBM's {\em business artifact model} 
pioneered in~\cite{Nigam03:artifacts}, successfully deployed in practice ~\cite{Bhatt-2005:Artifacts-pharm,Bhatt-2007:artifacts-customer-engagements,IGF-case-study:BPM-2009,Cordys-case-management,IBM-case-mgmt} and adopted in industrial standards.

In a nutshell, business artifacts (or simply ``artifacts'') model key business-relevant
entities, which are updated by a set of services that implement
business process tasks, specified declaratively by pre- and-post conditions. 
A collection of artifacts and services is called an {\em artifact system}. 
IBM has developed several variants of artifacts, of which the most recent is 
Guard-Stage-Milestone (GSM) \cite{Damaggio:BPM11,GSM:DEBS-2011}. 
The GSM approach provides rich structuring mechanisms for services, including 
parallelism, concurrency and hierarchy, and has been incorporated in the OMG standard 
for Case Management Model and Notation (CMMN) \cite{GSM-CMMN:2012,OMG:CMMN:Beta1}.

Artifact systems deployed in industrial settings typically specify complex workflows 
prone to costly bugs, whence the need for verification of critical properties.
Over the past few years, the verification problem for artifact systems has been intensively studied.
Rather than relying on general-purpose software verification tools suffering from well-known limitations,
the focus of the research community has been to identify practically relevant classes of artifact systems and 
properties for which {\em fully automatic} verification is possible.
This is an ambitious goal, since artifacts are infinite-state systems due to the presence of unbounded data. 
However, verification was shown to be decidable for significant classes of properties and artifact models. 

The present paper bridges the gap between the theory and practice 
of artifact verification by studying the implementation of a full-fledged and efficient artifact verifier.
The artifact model we verify is a variant of the Hierarchical Artifact System (HAS) model 
presented in \cite{pods16}.  In brief, a HAS consists of a database and a hierarchy (rooted tree) of {\em tasks}.
Each task has associated to it local evolving data consisting of a tuple of artifact variables and an updatable artifact relation.
It also has an associated set of {\em services}.
Each application of a service is guarded by a pre-condition on the database and local data and
causes an update of the local data, specified by a post condition (constraining the next artifact tuple)
and an insertion or retrieval of a tuple from the artifact relation. 
In addition, a task may invoke a child task with a tuple of parameters,
and receive back a result if the child task completes. A run of the artifact system is
obtained by any valid interleaving of concurrently running task services.
Properties of HAS are specified using an extension of Linear-Time Temporal logic (LTL).

In a previous study \cite{spinarxive}, we made a first attempt at implementing a verifier for a 
simple version of HAS using Spin \cite{spin}, the verification tool widely used
in the model checking community. However, as discussed in \cite{spinarxive}, 
Spin cannot handle some of the most useful 
features of artifacts which support unbounded data, such as {\em sets} of tuples 
(see Section \ref{sec:newmodel} for details).
Moreover, its performance is disappointing even after deploying a battery of non-trivial optimizations. 
This indicates the limited applicability of existing tools for HAS verification and suggests the need for
tailored approaches.

In this paper we present VERIFAS, an artifact verifier implementation built from scratch.
Our main contributions are the following.
\begin{itemize}\parskip=0in\itemsep=0in
\item We define HAS*, a novel variant of HAS which strikes a more practically relevant trade-off between expressivity and verification complexity, as demonstrated by its ability to specify a realistic set 
of business processes.  We adapt to HAS* the theory developed in \cite{pods16}, laying the groundwork for our implementation.
\item We implement VERIFAS, a fully automatic verifier for HAS*. The implementation makes crucial use of {\em novel optimization techniques}, with dramatic impact on performance. 
The optimizations are non-trivial and include 
concise symbolic representations, aggressive pruning in the search algorithm, and the use of 
highly efficient data structures.  
\item
We evaluate the performance of VERIFAS using both real-world and synthetic artifact systems 
and properties from a benchmark we create, bootstrapping
from existing sets of business process specifications and properties by extending them 
with data-aware features. To our knowledge, this is the first benchmark for business processes and properties that includes such data aware features. The experiments
highlight the impact of the various optimizations and parameters of both the artifact systems and properties. 
\item We adapt to HAS* a standard complexity measure of control flow used in software engineering,
{\em cyclomatic complexity} \cite{cyclomatic2}, and show experimentally, using the above benchmark, 
that cyclomatic complexity of HAS* specifications correlates meaningfully with verification times.
Since conventional wisdom in software engineering holds that well-designed, human readable programs 
have relatively low cyclomatic complexity, this is an indication that verification times 
are likely to be good for well-designed HAS* specifications. 
\end{itemize}

Taking this and other factors into account, the
experimental results show that our verifier performs very well on practically relevant
classes of artifact systems. Compared to the Spin-based verifier of \cite{spinarxive}, it not only 
applies to a much broader class of artifacts but also has a decisive performance advantage even on the simple artifacts 
the Spin-based verifier is able to handle.
To the best of our knowledge, this is the first implementation of practical significance of an artifact verifier 
with full support for unbounded data. 

The paper is organized as follows. We start by introducing in Section \ref{sec:newmodel} the
HAS* model supported by VERIFAS, and we
review LTL-FO, the temporal logic for specifying properties of HAS*.
Section \ref{sec:karp-miller} describes the implementation of VERIFAS by first reviewing in brief the theory 
developed of \cite{pods16}, particularly the symbolic representation technique used in establishing the theoretical results. 
We show an extension of the symbolic representation, called partial isomorphism type, to allow practical verification by adapting
the classic Karp-Miller algorithm \cite{karp-miller}. We then introduce three specialized optimizations 
to gain further performance improvement. We present our experimental results in Section \ref{sec:experiment}.
Finally, we discuss related work in Section \ref{sec:related} and conclude in Section \ref{sec:conclusion}. An appendix provides further technical details, our full running example, and a table of symbols.

%% file: newmodel.tex
\section{The Model}
\label{sec:newmodel}

In this section we present the variant of Hierarchical Artifact Systems used in our study.
The variant, denoted HAS*, differs from the HAS model used in \cite{pods16} in two respects.
On one hand, it {\em restricts} HAS as follows:
\vspace{-1mm}
\begin{itemize}\itemsep=0pt\parskip=0pt
\item it disallows arithmetic in service pre-and-post conditions
\item the underlying database schema uses an {\em acyclic} set of foreign keys
\end{itemize}
On the other hand, HAS* {\em extends} HAS by removing various restrictions:
\vspace{-1mm}
\begin{itemize}\itemsep=0pt\parskip=0pt
\item tasks may have multiple updatable artifact relations
\item each subtask of a given task may be called multiple times between task transitions
\item certain restrictions on how variables are passed as parameters among tasks, or inserted/retrieved from artifact relations,
are lifted
\end{itemize}

Because HAS* imposes some restrictions on HAS but removes others, it is incomparable to HAS.
Intuitively, the choice of HAS* over HAS as a target for verification is motivated by the fact that
HAS* achieves a more appealing trade-off between expressiveness and verification complexity.
The acyclic schema restriction, satisfied by the widely used Star (or Snowflake) schemas \cite{starschema1, starschema2}, is acceptable in return for the removal of various HAS restrictions limiting
modeling capability.  Indeed, as shown by our real-life examples, HAS* is powerful enough to model a wide variety of business processes.
While the current version of VERIFAS does not handle arithmetic,
the core verification algorithm can be augmented to include arithmetic
along the lines developed for HAS in \cite{pods16}. Limited use of aggregate functions can also be accommodated.
These enhancements are left for future work.

We now present the syntax and semantics of HAS*.
The formal definitions below are illustrated with an intuitive example of
the HAS* specification of a real-world order fulfillment business process
originally written in BPMN \cite{bpmn}.
The workflow allows customers to place orders and the supplier company to process the orders.
A detailed description of the example can be found in Appendix \ref{app:example}.

We begin by defining the underlying database schema.
\vspace{-1mm}
\begin{definition}
A \textbf{database schema} $\db$ is a finite set of relation symbols, where
each relation $R$ of $\db$ has an associated sequence of distinct attributes
containing the following:

\vspace{-2mm}
\begin{itemize}\itemsep=0pt\parskip=0pt
\item a key attribute $\emph{ID}$ (present in all relations),
\item a set of foreign key attributes $\{F_1, \dots, F_m\}$, and
\item a set of non-key attributes $\{A_1, \dots, A_n\}$ disjoint from \\
 $\{\emph{ID}, F_1, \dots, F_m\}$.
\end{itemize}
\vspace{-1mm}
To each foreign key attribute $F_i$ of $R$ is associated a relation $R_{F_i}$ of $\db$
and the inclusion dependency\footnote{The inclusion uses set semantics.} $R[F_i] \subseteq R_{F_i}[\emph{ID}]$.
It is said that $F_i$ references $R_{F_i}$.
\end{definition} \vspace{-1mm}

The assumption that the ID of each relation is a single attribute is made for simplicity, and multiple-attribute IDs
can be easily handled.

A database schema $\db$ is {\em acyclic} if there are no cycles in the references induced by foreign keys.
More precisely, consider the labeled graph FK whose nodes are the relations of the schema and
in which there is an edge from $R_i$ to $R_j$ labeled with $F$ if $R_i$
has a foreign key attribute $F$ referencing $R_j$.
The schema $\db$ is {\em acyclic} if the graph FK is acyclic.
All database schemas considered in this paper are acyclic.
\vspace*{-1mm}
\begin{example}
The order fulfillment workflow has the following database schema:
\vspace{-1mm}
\begin{itemize}\itemsep=0pt\parskip=0pt
\item
\dbcustomers $\mathtt{(\underline{\emph{ID}}, name, address, record)}$\\
\dbitems $\mathtt{(\underline{\emph{ID}}, item\_name, price)}$\\
\dbrecords $\mathtt{(\underline{\emph{ID}}, status)}$\\
\end{itemize}
\vspace{-5mm}

In the schema, the IDs are key attributes,
$\mathtt{price}$, $\mathtt{item\_name}$, $\mathtt{name}$, $\mathtt{address}$, $\mathtt{status}$
are non-key attributes, and $\mathtt{record}$ is a foreign key attribute
satisfying the dependency $\dbcustomers[record]$ $\subseteq \dbrecords[\emph{ID}]$.
Intuitively, the
$\dbcustomers$ table contains customer information with a foreign key pointing to
the customers' credit records stored in $\dbrecords$.
The $\dbitems$ table contains information on the items.
Note that the schema is acyclic as there is only one foreign key reference from $\dbcustomers$ to $\dbrecords$.
\end{example}
\vspace*{-1mm}

We assume two infinite, disjoint domains of IDs and data values, denoted by $\emph{DOM}_{id}$ and $\emph{DOM}_{val}$, and an additional
constant $\anull$ where $\anull \not\in \emph{DOM}_{id} \cup \emph{DOM}_{val}$
 ($\anull$ serves as a convenient default initialization value). 
The domain of all non-key attributes is  $\emph{DOM}_{val}$.
The domain of each key attribute ID of relation $R$ is an infinite subset $Dom(R.\emph{ID})$ of $\emph{DOM}_{id}$, and
$Dom(R.\emph{ID}) \cap Dom(R'.\emph{ID}) = \emptyset$ for $R \neq R'$.
The domain of a foreign key attribute $F$ referencing $R$ is $Dom(R.\emph{ID})$.
Intuitively, in such a database schema, each
tuple is an object with a \emph{globally} unique id. This id does not appear
anywhere else in the database except in foreign keys referencing it.
An {\em instance} of a database schema $\db$ is a mapping $D$ associating to each relation symbol $R$ a finite relation (set of tuples)
$D(R)$ of the same arity of $R$, whose tuples provide, for each attribute, a value from its domain. In addition,
$D$ satisfies all key and inclusion dependencies associated with the keys and foreign keys of the schema.
The active domain $D$, denoted $\texttt{adom}(D)$, consists of all elements of $D$.

We next proceed with the definition of tasks and services, described informally in the introduction.
Similarly to the database schema, we consider two infinite, disjoint sets $\varid$ of ID variables and $\varnum$ of
data variables. We associate to each variable $x$ its domain $Dom(x)$.
If $x \in \varid$, then $Dom(x) = \emph{DOM}_{id} \cup \{\anull\}$,
and if $x \in \varnum$, then $Dom(x) = \emph{DOM}_{val} \cup \{\anull\}$.
An {\em artifact variable} is a variable in $\varid \cup \varnum$.
If $\bar{x}$ is a sequence of artifact variables, a {\em valuation} of $\bar{x}$
is a mapping $\nu$ associating to each variable $x$ in $\bar{x}$ an element in $Dom(x)$.

\vspace{-1mm}
\begin{definition}
A \textbf{task schema} over database schema $\db$ is a tuple $T = \langle \bar{x}^T, \cals^T , \bar{x}^T_{in}, \bar{x}^T_{out} \rangle$ where
$\bar{x}^T$ is a sequence of artifact variables,
$\cals^T$ is a set of relation symbols not in $\db$, and
$\bar{x}^T_{in}$ and $\bar{x}^T_{out}$ are subsequences of $\bar{x}^T$.
For each relation $\cals \in \cals^T$,
we denote by $\mathtt{attr}(\cals)$ the set of attributes of $\cals$.
The domain of each variable $x \in \bar{x}^T$ and each attribute $A \in \mathtt{attr}(\cals)$
is either $\emph{DOM}_{val}  \cup \{\anull\}$ or $dom(R.ID) \cup \{\anull\}$ for some relation $R \in \db$.
In the latter case we say that the type of $x$ (or $A$) is  $\type(x) = R.ID$ ($\type(A) = R.ID$).
An \textbf{instance} $\rho$ of $T$ is a tuple $(\nu, S)$
where $\nu$ is a valuation of $\bar{x}^T$ and $S$ is an instance of $\cals^T$ such that
$S(\cals)$ is of the type of $\cals$ for each $\cals \in \cals^T$.
\end{definition}
\vspace{-1mm}

We refer to the relations in $\cals^T$ as the {\em artifact relations} of $T$ and to
$\bar{x}^T_{in}$ and $\bar{x}^T_{out}$ as the input and output variables of $T$.
We denote by $\bar x^T_{id} = \bar x^T \cap \varid$ and $\bar x^T_{\emph{val}} = \bar x^T \cap \varnum$.

\vspace{-1mm}
\begin{example}
The order fulfillment workflow has a task called \textbf{ProcessOrders}, which
stores the order data and processes the orders by interacting with other tasks.
It has the following artifact variables:
\vspace{-1.5mm}
\begin{itemize}\itemsep=0pt\parskip=0pt
\item ID variables: $\acustomerid$ of type $\dbcustomers.\emph{ID}$ and $\aitemid$ of type $\dbitems.\emph{ID}$
\item non-ID variables: $\astatus$ and $\ainstock$
\end{itemize}
\vspace{-1.5mm}
There are no input or output variables.
The task also has an artifact relation $\mathtt{ORDERS}(\acustomerid, \aitemid, \astatus, \ainstock)$
with attributes of the same types as the variables.
Intuitively, $\mathtt{ORDERS}$ stores the orders to be processed, where each order consists of
a customer and an ordered item. The variable $\astatus$ indicates the current status of the order
and $\ainstock$ indicates whether the item is currently in stock.
\end{example}
\vspace{-3mm}

We next define artifact schemas, essentially a hierarchy of task schemas with an underlying
database.
\begin{definition}
An \textbf{artifact schema} is a tuple $\mathcal{A} = \langle \calh, \db \rangle$
where $\mathcal{DB}$ is a database schema and $\calh$ is a rooted tree of task schemas over $\db$
with pairwise disjoint sets of artifact variables and distinct artifact relation symbols.
\end{definition}\vspace{-1mm}

The rooted tree $\calh$ defines the {\em task hierarchy}.
Suppose the set of tasks is $\{T_1, \ldots, T_k\}$.
For uniformity, we always take task $T_1$ to be the root of $\calh$.
We denote by $\preceq_\calh$ (or simply $\preceq$ when $\calh$ is understood) the
partial order on $\{T_1, \dots, T_k\}$ induced by $\calh$ (with $T_1$ the minimum).
For a node $T$ of $\calh$,
we denote by $\emph{tree(T)}$ the subtree of $\calh$ rooted at $T$,
$\emph{child}(T)$ the set of children of $T$ (also called {\em subtasks} of $T$),
$\emph{desc}(T)$ the set of descendants of $T$ (excluding $T$). Finally, $\emph{desc}^*(T)$ denotes
$\emph{desc}(T) \cup \{T\}$.
We denote by $\cals_\calh$
the relational schema $\cup_{1 \leq i \leq k} \cals^{T_i}$.
An instance of $\cals_\calh$ is a mapping associating to each $S \in \cals_\calh$ a finite relation of the same type.

\begin{figure}[!ht]
\vspace{-4mm}
\centering
\includegraphics[scale=0.5]{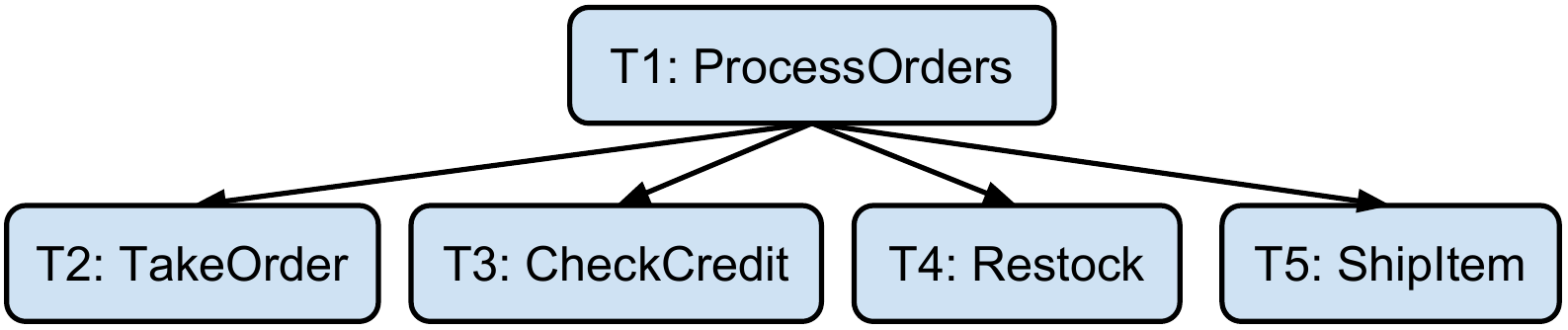}
\vspace{-2mm}
\caption{Tasks Hierarchy}
\label{fig:hierarchy-simple}
\vspace{-4mm}
\end{figure}

\vspace{-1mm}
\begin{example}
The order fulfillment workflow has 5 tasks: $T_1$: \textbf{ProcessOrders},
$T_2$:\textbf{TakeOrder}, $T_3$:\textbf{CheckCredit}, $T_4$: \textbf{Restock} and $T_5$:\textbf{ShipItem},
which form the hierarchy represented in Figure \ref{fig:hierarchy-simple}.
Intuitively, the root task \textbf{ProcessOrders} serves as a global coordinator which
maintains a pool of all orders and the child tasks \textbf{TakeOrder},
\textbf{CheckCredit}, \textbf{Restock} and \textbf{ShipItem} implement the 4 sequential stages in the fulfillment of an order.
At a high level, \textbf{ProcessOrders} repeatedly picks an order from its pool and
processes it with a stage by calling the corresponding child task.
After the child task returns, the order is either placed back into the pool or processed with the next stage.
For each order, the workflow first obtains the customer and item information using the \textbf{TakeOrder} task.
The credit record of the customer is checked by the \textbf{CheckCredit} task. If the record is good,
then \textbf{ShipItem} can be called to ship the item to the customer. If the requested item is unavailable,
then \textbf{Restock} must be called before \textbf{ShipItem} to procure the item.
\end{example}
\vspace{-1mm}

\vspace{-1mm}
\begin{definition}
An \textbf{instance} of an artifact schema $\mathcal{A} = \langle \calh, \db \rangle $ is a tuple
$I = \langle  \nu, stg, D,  S \rangle$ where
$D$ is a finite instance of $\mathcal{DB}$, $S$ a finite instance of $\cals_\calh$,
$\nu$ a valuation of
$\bigcup_{i=1}^k \bar{x}^{T_i}$,
and $stg$ (standing for ``stage'') a mapping of $\{T_1, \dots, T_k\}$ to
$\{\aactive, \aclosed \}$.
\end{definition}
\vspace{-1mm}

The stage $stg(T_i)$ of a task $T_i$ has the following intuitive meaning in the context of a run of its parent:
$\aactive$ says that $T_i$ has been called and has not yet returned its answer, and $\aclosed$ indicates that $T_i$
is not active.
A task $T_i$ can be called any number of times within a given run of its parent, but only
one instance of it can be active at any given time.

\vspace{-1mm}
\begin{example}
Figure \ref{fig:instance} shows a partial example of an instance of
the Order Fulfillment artifact system. The only active task is $\textbf{ProcessOrder}$.

\begin{figure}[!ht]
\vspace{-3mm}
\centering
\includegraphics[width=0.5 \textwidth]{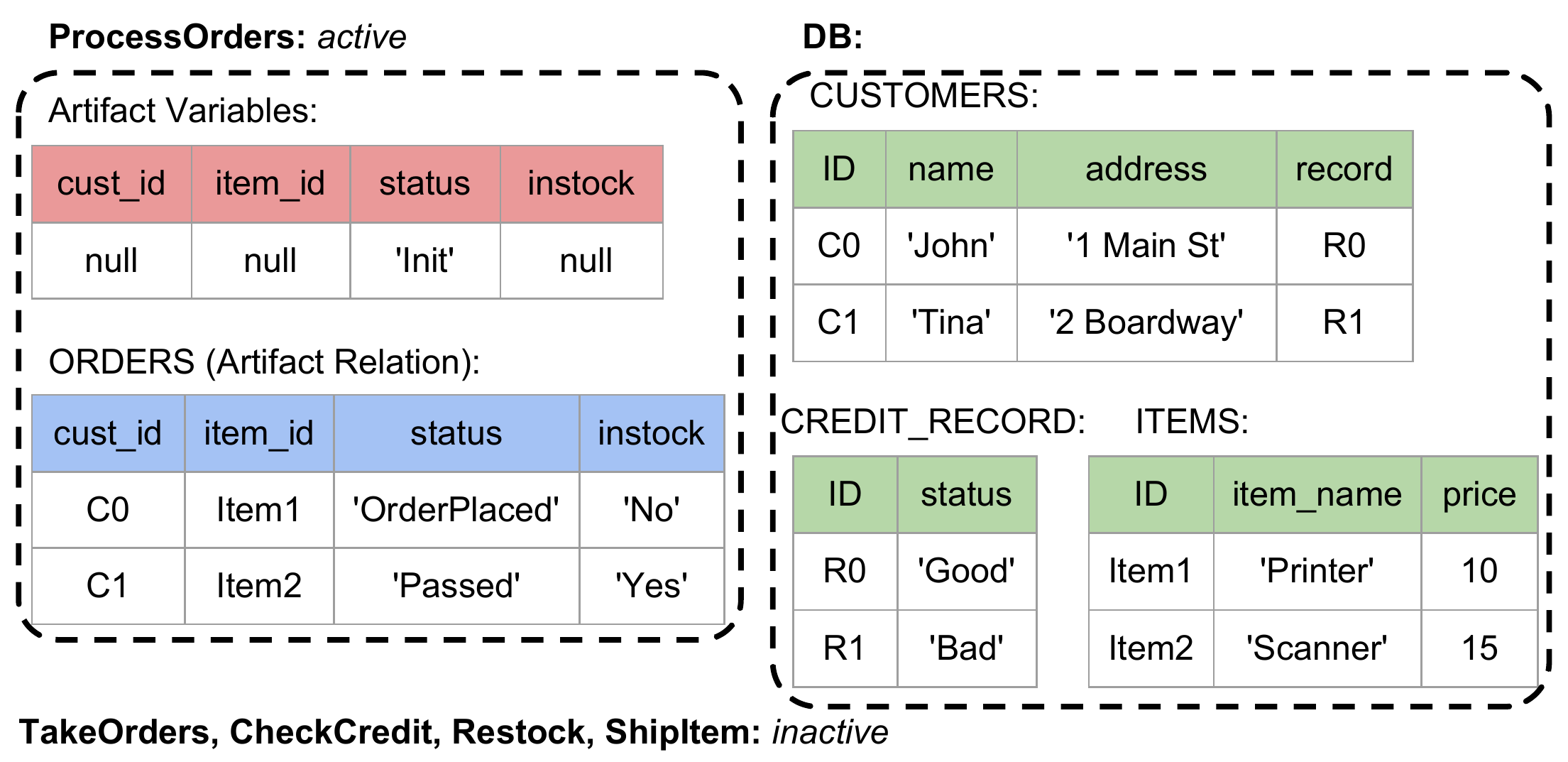}
\vspace{-4mm}
\caption{An instance of the Order Fulfillment workflow}
\label{fig:instance}
\vspace{-4mm}
\end{figure}

\end{example}
\vspace{-1mm}

For a given artifact schema $\mathcal{A} = \langle \calh, \mathcal{DB} \rangle$
and a sequence $\bar x$ of variables, a {\em condition} on $\bar x$ is a quantifier-free FO formula over
$\mathcal{DB} \cup \{=\}$ whose variables are included in $\bar{x}$.
The special constant $\anull$ can be used in equalities.
For each atom $R(x, y_1, \dots, y_m,$ $ z_1, \dots, z_n)$ of relation
$R(\emph{ID}, A_1, \dots, A_m, F_1, \dots, F_n) \in \db$,
$\{x, z_1, \dots, z_n\} \subseteq \varid$ and $\{y_1, \dots, y_m\} \subseteq \varnum$.
If $\alpha$ is a condition on $\bar x$,
$D$ an instance of $\mathcal{DB}$ and $\nu$ a valuation of $\bar x$,
we denote by $D \models \alpha(\nu)$ the fact that $D$ satisfies $\alpha$ with valuation $\nu$, with standard semantics.
For an atom $R(\bar{y})$ in $\alpha$ where $R \in \db$,
if $\nu(y) = \anull$ for some $y \in \bar{y}$, then $R(\nu(\bar{y}))$ is false (because $\anull$ does not occur in database relations).
Although conditions are quantifier-free,
$\exists$FO conditions can be easily simulated by adding variables to $\bar{x}^T$, so we use them as shorthand whenever convenient.

\vspace{-1mm}
\begin{example}
The $\exists$FO condition \\ 
$ \exists n \exists a \exists r \dbcustomers(\acustomerid, n, a, r) \land \dbrecords(r, \text{``Good''})$
states that the customer with ID $\acustomerid$ has good credit.
\end{example}
\vspace{-1mm}


We next define services of tasks. We start with internal services, which update
the artifact variables and artifact relation of the task.
Intuitively, internal services implement local actions, that do not involve any other task.

\vspace{-2mm}
\begin{definition}
Let $T = \langle \bar{x}^T, \cals^T, \bar{x}^T_{in}, \bar{x}^T_{out} \rangle$
be a task of an artifact schema $\mathcal{A}$.
An \textbf{internal service} $\sigma$ of $T$
is a tuple $\langle \pi, \psi, \bar y, \delta \rangle$ where:
\vspace*{-1.5mm}
\begin{itemize}\itemsep=0pt\parskip=0pt
\item $\pi$ and $\psi$, called \emph{pre-condition} and \emph{post-condition},
respectively, are conditions over $\bar{x}^T$
\item $\bar y$ is the set of {\em propagated variables}, where $\bar{x}^T_{in} \subseteq \bar{y} \subseteq  \bar{x}^T$;
\item $\delta$, called the {\em update}, is a subset of $\{+\cals_i(\bar{z}), -\cals_i(\bar{z}) | \cals_i \in \cals^T,
\bar{z} \subseteq \bar{x}^T, \type(\bar{z}) = \type(\mathtt{attr}(\cals_i)) \}$ of size at most 1.
\item if $\delta \neq \emptyset$ then  $\bar{y} = \bar{x}^T_{in}$
\end{itemize}
\end{definition}
\vspace{-2mm}

Intuitively, an internal service $\sigma$ of $T$ can be called only when
the current instance satisfies the pre-condition $\pi$.
The update on variables $\bar{x}^T$ is valid if the next instance satisfies
the post-condition $\psi$ and the values of propagate variables $\bar{y}$ stay unchanged.

Any task variable that is not propagated can be changed arbitrarily during a task activation,
as long as the post condition holds.
This feature allows services to also model actions by external actors who provide
input into the workflow by setting the value of non-propagated variables.
Such actors may even include humans or other parties whose behavior is not deterministic.
For example, a bank manager carrying out a ``loan decision'' action
can be modeled by a service whose result is stored in a non-propagated variable and whose value is
restricted by the post-condition to either ``Approve'' or ``Deny''. Note that deterministic actors are modeled
by simply using tighter post-conditions.

When $\delta = \{+\cals_i(\bar{z})\}$, a tuple containing the {\em current} value of
$\bar{z}$ is inserted into $\cals_i$.
When $\delta = \{-\cals_i(\bar{z})\}$, a tuple is chosen 
and removed from $\cals_i$ and the {\em next} value of $\bar{z}$
is assigned with the value of the tuple.
Note that $\bar{x}^T_{in}$ are always propagated, and no other variables are propagated if $\delta \neq \emptyset$.
The restriction on updates and variable propagation may at first appear mysterious.
Its underlying motivation is that allowing simultaneous artifact relation updates and variable propagation
turns out to raise difficulties for verification, while
the real examples we have encountered do not require this capability.

\vspace{-1mm}
\begin{example}
The \textbf{ProcessOrders}
task has 3 internal services: \emph{Initialize}, \emph{StoreOrder} and \emph{RetrieveOrder}.
Intuitively, \emph{Initialize} creates a new order with $\acustomerid = \aitemid = \anull$.
When \emph{RetrieveOrder} is called, an order is chosen non-deterministically and removed
from $\mathtt{ORDERS}$ for processing, and $(\acustomerid, \aitemid, \\ \astatus, \ainstock)$ is set to be the chosen tuple.
When \emph{StoreOrder} is called, the current order $(\acustomerid, \aitemid, \astatus, \ainstock)$ is inserted into
$\mathtt{ORDERS}$. The latter two services are specified as follows.

\vspace{2mm}
\noindent
\emph{RetrieveOrder}: \\
Pre: $\acustomerid = \anull \land \aitemid = \anull$ \\
Post: $\mathtt{True}$ \\
Update: $\{-\mathtt{ORDERS}(\acustomerid, \aitemid, \astatus, \ainstock)\}$

\vspace{2mm}
\noindent
\emph{StoreOrder}: \\
Pre: $\acustomerid \neq \anull \land \aitemid \neq \anull \land \astatus \neq \text{``Failed"}$ \\
Post: $\acustomerid = \anull \land \aitemid = \anull \land \astatus = \text{``Init"}$ \\
Update: $\{+\mathtt{ORDERS}(\acustomerid, \aitemid, \astatus, \ainstock)\}$

\vspace{1mm}
\noindent
The sets of propagated variables are empty for both services.

\end{example}
\vspace{-1mm}

\begin{figure}[!ht]
\vspace{-2mm}
\centering
\includegraphics[width=0.5 \textwidth]{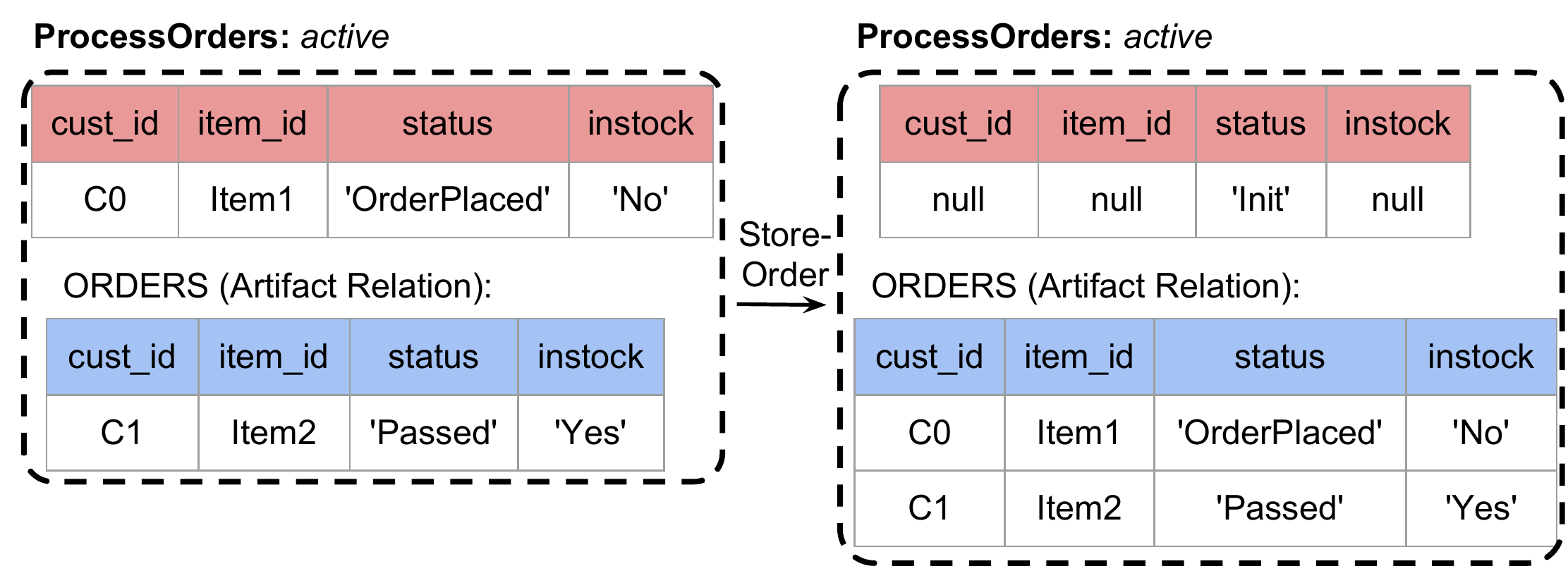}
\vspace{-5mm}
\caption{Transition caused by an internal service}
\label{fig:has-transition}
\vspace{-4mm}
\end{figure}

An internal service of a task $T$ specifies transitions that
modify the variables $\bar{x}^T$ of $T$ and the contents of $\cals^T$.
Figure \ref{fig:has-transition} shows an example of a transition that results from applying the service
\emph{StoreOrder} of the \textbf{ProcessOrders} task.

As seen above, internal services of a task cause transitions on the data local to the task.
Interactions among tasks are specified using two kinds of special services,
called the \emph{opening-services} and \emph{closing-services}.
Specifically, each task $T$ is equipped with an opening service $\sigma_{T}^o$ and a closing service
$\sigma_{T}^c$. Each non-root task $T$ can be activated by its parent task via a call to $\sigma_{T}^o$
which includes passing parameters to $T$  that initialize its input variables $\bar{x}_{in}^{T}$.
When $T$ terminates (if ever), it returns to the parent the contents of its output variables
$\bar{x}_{out}^{T}$ via a call to $\sigma_{T}^c$. Moreover, calls to $\sigma_{T}^o$ are guarded by a condition on the
parent's artifact variables, and closing calls to $\sigma_{T}^c$ are guarded by a condition on the artifact variables of $T$.
The formal definition is provided in Appendix \ref{app:semantics}.

For uniformity of notation, we also equip the root task $T_1$ with a service $\sigma^o_{T_1}$ with pre-condition {\em true}
and a service $\sigma^c_{T_1}$ whose pre-condition is \emph{false}
(so it never occurs in a run).
For a task $T$ we denote by $\Sigma_T$ the set of its internal services,
$\Sigma_T^{oc} = \Sigma_T \cup \{\sigma^o_T, \sigma^c_T\}$, and
$\Sigma_T^{\emph{obs}} = \Sigma_T^{oc} \cup \{\sigma^o_{T_c}, \sigma^c_{T_c} \mid T_c \in \emph{child}(T)\}$.
Intuitively, $\Sigma_T^{\emph{obs}}$ consists of the services observable locally in runs of task $T$.

\begin{figure}[!ht]
\vspace{-3mm}
\centering
\includegraphics[width=0.5 \textwidth]{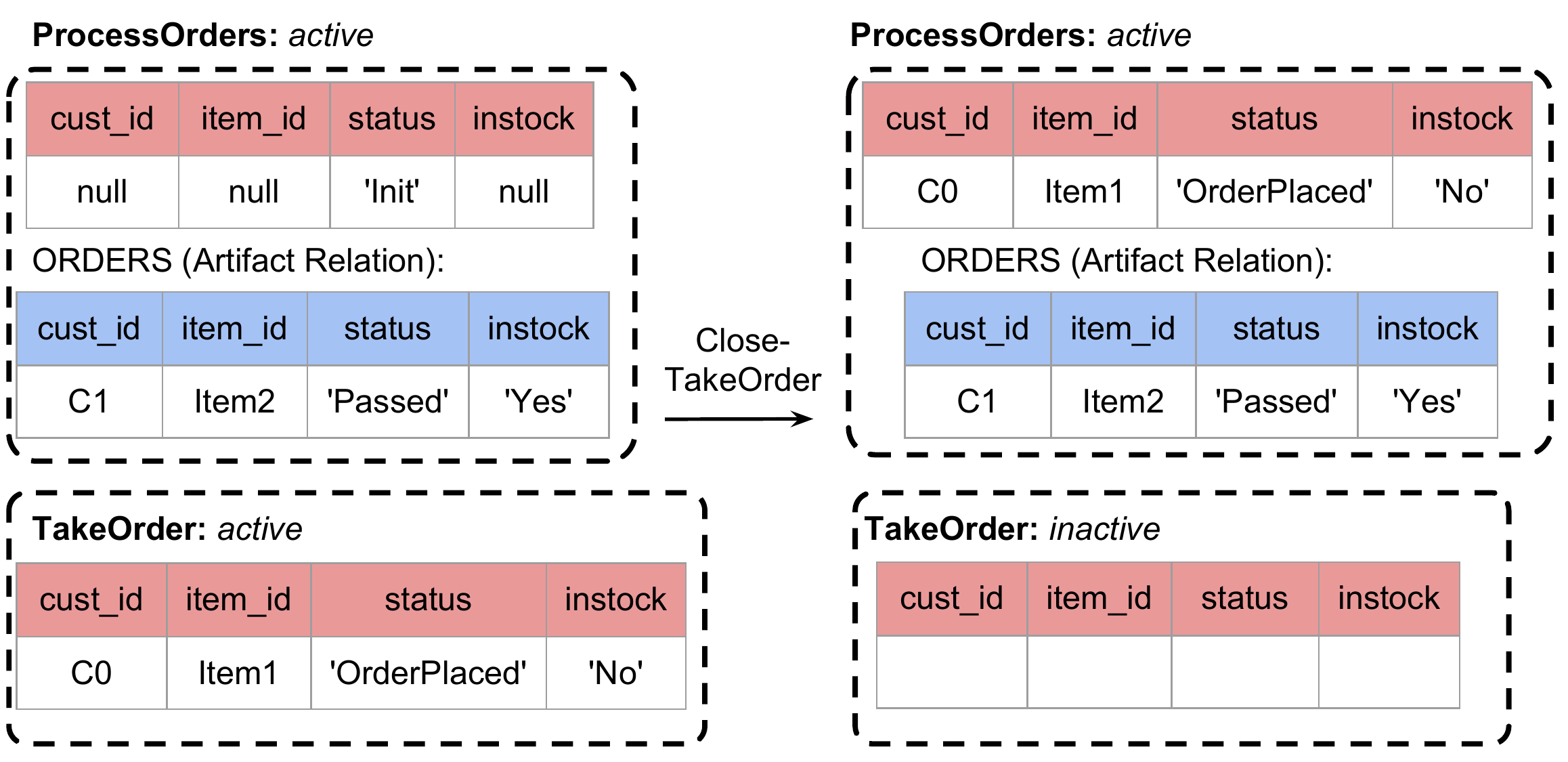}
\vspace{-4mm}
\caption{Transition with a Closing Service}
\label{fig:fig-closing}
\vspace{-4mm}
\end{figure}

\vspace{-1mm}
\begin{example}
As the root task, the opening condition of \textbf{ProcessOrders}
is $\mathtt{True}$ and closing condition is $\mathtt{False}$.
All variables are initialized to $\anull$.

The opening condition of \textbf{TakeOrder} is $\astatus = \text{``Init''}$ in
task \textbf{ProcessOrders}, meaning that the customer and item information
have not yet been entered by the customer.
The task contains $\acustomerid$, $\aitemid$, $\astatus$ and $\ainstock$ as variables
(with no input variable). When this task is called, the customer enters
the information of the order ($\acustomerid$ and $\aitemid$) and
the status of the order is set to $\text{``OrderPlaced"}$.
An external service determines whether the item is in stock or not and
sets the value of $\ainstock$ accordingly.
All variables are output variables returned to the parent task.
The closing condition is $\acustomerid \neq \anull \land \aitemid \neq \anull$.
When it holds, \textbf{TakeOrder} can be closed,
and the values of these variables are passed to \textbf{ProcessOrders} (to the variables
with the same names\footnote{While the formal definition disallows using the same variable names in different tasks,
we do so for convenience, since the variable names can be easily disambiguated using the task name.}).
Figure \ref{fig:fig-closing} illustrates a transition
caused by the closing service of \textbf{TakeOrder}.
\end{example}
\vspace{-1mm}

We are finally ready to define HAS*.

\vspace{-1mm}
\begin{definition}
A \emph{Hierarchical Artifact System}* (HAS*) is a triple $\Gamma = \langle \mathcal{A}, \Sigma, \Pi \rangle$,
where $\mathcal{A}$ is an artifact schema, $\Sigma$ is a set of services over $\mathcal{A}$ including
$\sigma^o_T$ and  $\sigma^c_T$ for each task $T$ of $\mathcal{A}$,
and $\Pi$ is a condition over $\bar x^{T_1}$ (the global pre-condition of $\Gamma$),
where $T_1$ is the root task.
\end{definition}
\vspace*{-1mm}


We next define the semantics of HAS*. Intuitively, a run of a HAS* on a database $D$ consists of
an infinite sequence of transitions among HAS* instances (also referred to as configurations, or snapshots),
starting from an initial artifact tuple satisfying pre-condition $\Pi$, and empty artifact relations.
The intuition is that at each snapshot,
a transition can be made at an active task $T$ by applying either an internal service of $T$,
the opening service of an inactive subtask $T_c$, or the closing service of $T$.
In addition, we require that an internal service of $T$ can only be applied
after all active subtasks of $T$ have returned their answer.
Given two instances $I$, $I'$ and a service $\sigma$,
we denote by $I \goto{\sigma} I'$ if there is a valid transition from $I$ to $I'$ by applying $\sigma$.
The full definition of transitions can be found in Appendix \ref{app:semantics}.

We next define {\em runs} of artifact systems. We will assume that runs are {\em fair},
i.e. no task is starved forever by other running tasks. Fairness is commonly ensured by schedulers in multi-process systems.
We also assume that runs are non-blocking, i.e. for each task that has not yet returned its answer,
there is a service applicable to it or to one of its descendants.

\vspace{-1mm}
\begin{definition}
Let $\Gamma = \langle \mathcal{A}, \Sigma, \Pi \rangle$ be an artifact system,
where $\mathcal{A} = \langle \calh, \mathcal{DB} \rangle$.
A \emph{run} of $\Gamma$ on database instance $D$ over $\mathcal{DB}$
is an infinite sequence $\rho = \{ (I_i, \sigma_i) \}_{i \geq 0}$,
where each $I_i$ is an instance $(\nu_i, stg_i, D, S_i)$ of $\cala$, $\sigma_i \in \Sigma$,
$\sigma_0 = \sigma^o_{T_1}$, $D \models \Pi(\nu_0)$, $stg_0 = \{T_1 \mapsto \aactive, T_i \mapsto \aclosed \mid 2 \leq i \leq k\}$,
$S_0 = \{\cals_{\calh} \mapsto \emptyset \}$, and for each $i > 0$
$I_{i-1} \goto{\sigma_{i}} I_{i}$. In addition,
for each $i \geq 0$ and task $T$ active in $I_i$, there exists $j > i$ such that
$\sigma_j \in \bigcup_{T' \in \emph{desc}^*(T)} \Sigma^{oc}_{T'}$.
\end{definition}
\vspace{-1mm}

We denote by $Runs(\Gamma)$ the set of runs of $\Gamma$.
Observe that all runs of $\Gamma$  are infinite.
In a given run, the root task itself may have an infinite run,
or other tasks may have infinite runs.
However, if a task $T$ has an infinite run,
then none of its ancestor tasks can make an internal transition or return
(although they can still call other children tasks).

Because of the hierarchical structure of HAS*, and the locality of task specifications,
the actions of independent tasks running concurrently can be arbitrarily interleaved.
In order to express properties of HAS* in an intuitive manner, it will be useful to ignore such interleavings and
focus on the {\em local runs} of each task, consisting of the transitions affecting the local variables and artifact
relations of the task, as well as interactions with its children tasks. A local run of $T$ induced by  $\rho$ is a subsequence $\rho_T$ of $\rho$
  corresponding to transitions caused by $T$'s observable services (call these {\em observable $T$-transitions}).
  $\rho_T$ starts from an opening service $T$-transition and includes all subsequent observable $T$-transitions up to
  the first occurrence of a closing service $T$-transition (if any). See Appendix~\ref{app:semantics} for the formal definition.
We denote by $Runs_T(\rho)$ the set of local runs of $T$ induced by the run $\rho$ of $\Gamma$, and
$Runs_T(\Gamma) = \bigcup_{\rho \in Runs(\Gamma)}~ Runs_T(\rho)$.

\vspace{-1mm}
\subsection{Specifying properties of artifact systems}

In this paper we focus on verifying temporal properties of local runs of tasks in an artifact system.
For instance, in a task implementing an e-commerce application, we would like to specify properties such as:
\vspace{-1mm}
\begin{itemize}
\item[] $(\dag)$ If an order is taken and the ordered item is out of stock,
then the item must be restocked before it is shipped.
\end{itemize}
\vspace{-1mm}

In order to specify such temporal properties we use, as in previous work, an extension of LTL (linear-time temporal logic).
LTL is propositional logic augmented with temporal operators such as {\bf G} (always),
{\bf F} (eventually), {\bf X} (next) and {\bf U} (until)
(e.g., see \cite{ltl}).  An LTL formula $\varphi$ with propositions $prop(\varphi)$ defines a property of sequences
of truth assignments to $prop(\varphi)$.  For example, ${\bf G}p$ says that $p$ always holds in the sequence, {\bf F}$p$
says that $p$ will eventually hold, $p {\bf U} q$ says that $p$ holds at least until $q$ holds,
and ${\bf G}(p \rightarrow {\bf F}q)$ says that whenever $p$ holds,
$q$ must hold subsequently.

An LTL-FO property\footnote{The variant of LTL-FO used here differs from some previously defined in that the FO formulas
interpreting propositions are quantifier-free. By slight abuse we use here the same name.}  of a task $T$
is obtained starting from an LTL formula using some set $P \cup \Sigma_T^{obs}$ of propositions.
Propositions in $P$ are interpreted as conditions over the variables $\bar{x}^T$
of $T$ together with some additional \emph{global} variables $\bar y$,
shared by different conditions and allowing to connect the states of the task at different moments in time.
The global variables are universally quantified over the entire property.
Recall that $\Sigma_T^{obs}$ consists of the services observable in local runs of $T$ (including calls and returns from
children tasks). A proposition $\sigma \in \Sigma_T^{obs}$ indicates the application of service $\sigma$ in a given transition.

LTL-FO is formally defined in Appendix~\ref{app:semantics}. We provide a flavor thereof using the example property $(\dag)$.
The property is of the form
$\varphi = {\bf G} (p \rightarrow (\neg q \ {\bf U } \ r))$, which means if $p$ happens,
then in what follows, $q$ will not happen until $r$ is true. Here $p$ says that
the \textbf{TakeOrder} task returned with an out-of-stock item, $q$ states that
the \textbf{ShipItem} task is called with the same item, and $r$ states that
the service \textbf{Restock} is called to restock the item.
Since the item mentioned in $p$, $q$ and $r$ must be the same,
the formula requires using a global variable $i$ to record the item ID.
This yields the following LTL-FO property: 
\vspace{-1mm}
\begin{align*}
\forall i \ & \mathbf{G} (( \sigma^c_{\mathtt{TakeOrder}} \land \aitemid = i \land \ainstock = \text{``No''} ) \rightarrow \\
            & ( \neg (\sigma^o_{\mathtt{ShipItem}} \land \aitemid = i) {\ \bf U \ } (\sigma^o_{\mathtt{Restock}} \land \aitemid = i)  ))
\end{align*}
\vspace{-1mm}

A correct specification can enforce $(\dag)$ simply
by requiring in the pre-condition of $\sigma^o_{\mathtt{ShipItem}}$ that the item is in stock.
One such pre-condition is $(\ainstock=\text{``Yes''} \land \astatus=\text{``Passed''})$,
meaning that the item is in stock and the customer passed the credit check.
However, in a similar specification where the $\ainstock=\text{``Yes''}$
test is performed within \textbf{ShipItem} (i.e. in the pre-conditions of all shipping internal services)
instead of the opening service of \textbf{ShipItem},
the LTL-FO property $(\dag)$ is violated because \textbf{ShipItem} can be opened without first calling
the \textbf{Restock} task.
Our verifier would detect this error and produce a counter-example illustrating the violation.




We say that a local run $\rho_T$ of task $T$ satisfies
$\forall \bar y \varphi_f$, where $prop(\varphi) = P \cup \Sigma_T^{obs}$,
if $\varphi$ is satisfied,  for all valuations of $\bar y$ in
$DOM_{id} \cup DOM_{val} \cup \{\anull\}$, by the sequence of truth assignments to $P \cup \Sigma_T^{obs}$ induced by $f$
on $\rho_T$.
More precisely, let $(I_i,\sigma_i)$ denote the $i^{th}$ snapshot of $\rho_T$.
For each $p \in P$, the truth value induced for $p$ in $(I_i,\sigma_i)$ is the truth value of the condition $f(p)$
in $I_i$; a proposition $\sigma \in \Sigma_T^{obs}$ holds in $(I_i,\sigma_i)$ if $\sigma_i = \sigma$.
A task $T$ satisfies $\forall \bar y \varphi_f$
if $\rho_T$ satisfies $\forall \bar y \varphi_f$ for every $\rho_T \in \emph{Runs}_T(\Gamma)$.
Note that the database is fixed for each run, but may be different for different runs.

A classical result in model checking states that for every LTL formula $\varphi$, one can construct a finite-state
automaton $B_\varphi$, called a B\"{u}chi automaton, that
accepts precisely the infinite sequences of truth assignments to $prop(\varphi)$ that satisfy $\varphi$. A B\"{u}chi automaton is syntactically just a finite-state automaton,
which accepts an infinite word if it goes
infinitely often through an accepting state \cite{VW:LICS:86,SVW87}.
Here we are interested in evaluating LTL-FO formulas $\forall \bar y \varphi_f$ on both infinite {\em and} finite runs (infinite runs occur
when a task runs forever).  It is easily seen that for the $B_\varphi$ obtained by the standard construction
there is a subset $Q^{\emph{fin}}$ of its states such that $B_\varphi$
viewed as a classical finite-state automaton with final states $Q^{\emph{fin}}$
accepts precisely the finite words that satisfy $\varphi$.

\vspace{-1mm}
\begin{remark}
In \cite{pods16} we consider a more complex logic for specifying properties of artifact systems, called Hierarchical LTL-FO (HLTL-FO).
Intuitively, an HLTL-FO formula uses as building blocks LTL-FO formulas as above, acting on local runs of individual tasks,
but can additionally recursively state HLTL-FO properties on runs resulting from calls to children tasks.
As shown in \cite{pods16}, verification of HLTL-FO properties can be reduced to satisfiability of LTL-FO properties
by individual tasks. Our implementation focuses on verification of LTL-FO properties of individual
tasks. While this could be used as a building block for verifying complex HLTL-FO properties, verification of
LTL-FO properties of individual tasks is in fact adequate in most practical situations we have encountered.
\end{remark}
\vspace{-1mm}

%% file: karp-miller.tex
\newcommand{\asucc}{\ensuremath{\mathtt{succ}}}
\newcommand{\aconj}{\ensuremath{\mathtt{conj}}}
\newcommand{\aflat}{\ensuremath{\mathtt{flat}}}
\newcommand{\apos}{\ensuremath{\mathtt{pos}}}
\newcommand{\aeqc}{\ensuremath{\mathtt{eq}}}
\newcommand{\auneql}{\ensuremath{\mathtt{ue}}}
\newcommand{\imax}{\ensuremath{\cali_{\mathtt{max}}}}
\newcommand{\irep}{\ensuremath{\cali_{\mathtt{rep}}}}

\section{VERIFAS} \label{sec:karp-miller}

\newcommand{\tauin}{\tau_{in}}
\newcommand{\tauout}{\tau_{out}}
\newcommand{\nuin}{\nu_{in}}
\newcommand{\nuout}{\nu_{out}}
\newcommand{\II}{\{(I_i,\sigma_i)\}_{0 \leq i < \gamma}}
\newcommand{\IIw}{\{(I_i,\sigma_i)\}_{0 \leq i < \omega}}
\newcommand{\RS}{\{(\rho_i, \sigma_i)\}_{0 \leq i < \gamma}}
\newcommand{\RSw}{\{(\rho_i, \sigma_i)\}_{0 \leq i < \omega}}
\newcommand{\trt}{\tilde{\rho}_T}
\newcommand{\trtc}{\tilde{\rho}_{T_c}}
\newcommand{\trtp}{\tilde{\rho}_{T_p}}
\newcommand{\ID}{\emph{ID}}

\newcommand{\Ret}{\ensuremath{\texttt{Retrieve} } }
\newcommand{\Prop}{\ensuremath{\texttt{Prop} } }
\newcommand{\merge}{\ensuremath{\texttt{merge} } }
\newcommand{\Add}{\ensuremath{\texttt{Add} } }
\newcommand{\reach}{\ensuremath{\texttt{Reach} } }
\newcommand{\adom}{\ensuremath{\texttt{adom} } }
\newcommand{\vmap}{\ensuremath{\texttt{vmap} } }
\newcommand{\connect}{\ensuremath{\texttt{connect} } }

\newcommand{\e}{\epsilon}
\newcommand{\GP}{\emph{GP}}
\newcommand{\hexp}{\text{-}\exp}

In this section we describe the implementation of VERIFAS.
We begin with a brief review of the theory developed in \cite{pods16} that is relevant to the implementation.

\subsection{Review of the Theory} \label{sec:verification}

The decidability and complexity results of \cite{pods16} can be extended to HAS* by 
adapting the proofs and techniques developed there.  We can show the following.
\vspace*{-1mm}
\begin{theorem} \label{thm:has*}
Given a HAS* $\Gamma$ and an LTL-FO formula $\varphi$ for a task $T$ in $\Gamma$,
it is decidable in {\sc expspace} whether $\Gamma$ satisfies $\varphi$.
\end{theorem}
\vspace*{-1mm}
We outline informally the roadmap to verification developed in \cite{pods16}, which is the starting point
for the implementation.
Let $\Gamma$ be a HAS* and $\varphi$ an LTL-FO formula for some task $T$ of $\Gamma$.
We would like to verify that every local run of $T$ satisfies $\varphi$.
Since there are generally infinitely many such local runs due so the unbounded data domain, and each run can be infinite,
an exhaustive search is impossible.  This problem is addressed in \cite{pods16} by developing a symbolic representation
of local runs.  Intuitively, the symbolic representation has two main components:
\vspace*{-1.5mm}
\begin{itemize}\itemsep=0pt\parskip=0pt
\item [(i)] the {\em isomorphism type} of the artifact variables, describing symbolically the structure of the portion of the database 
reachable from the variables by navigating foreign keys
\item  [(ii)] for each artifact relation and isomorphism type, the number of tuples in the relation that share that isomorphism type
\end{itemize}
\vspace*{-1.5mm}
Observe that because of (ii), the symbolic representation is not finite state.  
Indeed, (ii) requires maintaining a set of counters, which can grow unboundedly. 

The heart of the proof in \cite{pods16} is showing that it is sufficient to verify symbolic runs rather than actual runs.
That is, for every LTL-FO formula $\varphi$,
all local run of $T$ satisfy $\varphi$ iff all symbolic local runs of $T$ satisfy $\varphi$.
Then the verification algorithm checks that there is no symbolic local run of $T$ violating $\varphi$ (so satisfying $\neg\varphi$).
The algorithm relies on a reduction to (repeated\footnote{Repeated reachability is needed for infinite runs.}) state reachability in Vector Addition Systems with States (VASS) \cite{vass}.
Intuitively, VASS are finite-state automata augmented with non-negative counters that can be incremented and decremented (but not tested for zero).
This turns out to be sufficient to capture the information described above. 
The states of the VASS correspond to the isomorphism types of the artifact variables, combined with states of the 
B\"{u}chi automaton needed to check satisfaction of $\neg\varphi$.

%

The above approach can be viewed as symbolically running the HAS* specification.
Consider the example in Section \ref{sec:newmodel}. 
After the \textbf{TakeOrder} task is called and returned, 
one possible local run of \textbf{ProcessOrders} might impose a set of constraints 
$\{\aitemid \neq \anull, \acustomerid \neq \anull, \astatus = \text{``OrderPlaced''}, \ainstock = \text{``Yes''}\}$ 
onto the artifact tuple of \textbf{ProcessOrders}.
Now suppose the \textbf{CheckCredit} task is called. 
The local run can make the choice that the customer has good credit.
Then when \textbf{CheckCredit} returns, the above set of constraints is updated with constraint \\
$\{\acustomerid.\arecord.\astatus = \text{``Good''}\}$, which means that in the read-only database,
the credit record referenced by $\acustomerid$ via foreign key satisfies $\astatus = \text{``Good''}$.
Next, suppose the \emph{StoreOrder} service is applied in \textbf{ProcessOrders}. 
Then we symbolically store the current set of constraints by 
increasing its corresponding counter by 1. The set of constraints of the artifact tuple is reset to
$\{\aitemid = \anull, \acustomerid = \anull, \astatus = \text{``Init''}\}$
as specified in the post-condition of \emph{StoreOrder}. 

Although decidability of verification can be shown as outlined above, 
implementation of an efficient verifier is challenging.
The algorithm that directly translates the artifact specification and the LTL-FO property into 
VASS's and checks (repeated) reachability is impractical because
the resulting VASS can have exponentially many states and counters in the input size,
and state-of-the-art VASS tools can only handle a small number of counters ($<$100) \cite{mist}.
To mitigate the inefficiency, VERIFAS never generates the whole VASS but 
instead lazily computes the symbolic representations on-the-fly. Thus, it only generates {\em reachable} symbolic states,
whose number is usually much smaller. In addition, isomorphism types in the symbolic representation are replaced by
\emph{partial isomorphism types}, which store only the subset of constraints on the variables
imposed by the current run, leaving the rest unspecified. 
This representation is not only more compact, but also results in a significantly smaller search space in practice.

In the rest of the section, we first introduce our revised symbolic representation based on partial isomorphism types. 
Next, we review the classic Karp-Miller algorithm adapted to the symbolic version of HAS* 
for solving state reachability problems. 
Three specialized optimizations are introduced to improve the performance. 
In addition, we show that our algorithm with the optimizations can be extended to solve the repeated state reachability problems
so that full LTL-FO verification of infinite runs can be carried out.
For clarity, the exposition in this section focuses on specifications with a single task.
The actual implementation extends these techniques to the full model with arbitrary number of tasks.

%


\subsection{Partial Isomorphism Types}
\label{sec:partial}

We start with our symbolic representation of local runs with partial isomorphism types.
Intuitively, a partial isomorphism type captures the necessary constraints imposed by the current run 
on the current artifact tuple and the read-only database. 
We start by defining \emph{expressions}, which denote variables, constants and 
navigation via foreign keys from id variables or attributes. An expression is either:
\vspace*{-2mm}
\begin{itemize}\itemsep=0pt\parskip=0pt
\item a constant $c$ occurring in $\Gamma$ or $\varphi$, or
\item a sequence $\xi_1.\xi_2.\ldots\xi_m$, where $\xi_1$ is an id artifact variable $x$ or
an id attribute $A$ of some artifact relation $\cals$, $\xi_2$ is an attribute of $R \in \db$ where $R.\emph{ID} = \type(\xi_1)$, 
and for each $i$, $2 \leq i < m$,
$\xi_i$ is a foreign key and $\xi_{i+1}$ is an attribute in the relation referenced by $\xi_i$.
\end{itemize}
\vspace*{-1.5mm}
We denote by $\cale$ the set of all expressions. 
Note that the length of expressions is bounded because of the acyclicity of the foreign keys, so $\cale$ is finite.


We can now define partial isomorphism types.
%
%

\vspace*{-2mm}
\begin{definition} \label{def:partial-isomorphism-type}
A \textbf{partial isomorphism type} $\tau$ is an undirected graph over $\cale$ with each edge labeled by $=$ or $\neq$, such that
the equivalence relation $\sim$ over $\cale$ induced by the edges labeled with $=$ satisfies:
\vspace*{-1mm}
\begin{enumerate}\itemsep=0pt\parskip=0pt
\item for every $e, e' \in \cale$ and every attribute $A$, 
if $e \sim e'$ and $\{e.A, \\ e'.A\} \subseteq \cale$ then $e.A \sim e'.A$, and
\item $(e_1, e_2, \neq) \in \tau$ implies that $e_1 \not\sim e_2$ and for every $e_1' \sim e_1$ and $e_2' \sim e_2$,
$(e_1', e_2', \neq) \in \tau$.
\end{enumerate}
\end{definition}
\vspace*{-2mm}

Intuitively, a partial isomorphism type keeps track of a set of ``$=$'' and ``$\neq$'' constraints and their implications
among $\cale$.
Condition 1 guarantees satisfaction of the key and foreign key dependencies.
Condition 2 guarantees that there is no contradiction among the $\neq$-edges and the $=$-edges.
In addition, the connection between two expressions can also be ``unknown'' if they are not connected by an edge.
The full isomorphism type can be viewed as a special case of partial isomorphism type where the undirected graph is complete.
In the worst case, the total number of partial isomorphism types is no smaller than the number of full isomorphism types
so using partial isomorphism types does not improve the complexity upper bound. 
In practice, however, since the number of constraints imposed by a run
is likely to be small, using partial isomorphism types can greatly reduce the search space.

\begin{figure}[!ht]
\centering
\includegraphics[width=0.48\textwidth]{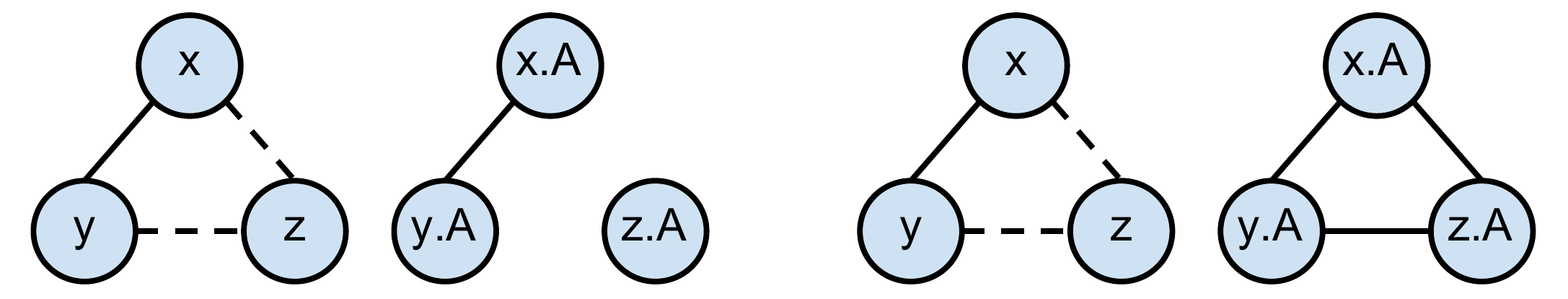}
\vspace*{-6mm}
\caption{Partial and Full Isomorphism Types}
\vspace*{-3mm}
\label{fig:partial}
\end{figure}

\vspace*{-2mm}
\begin{example} \label{exm:partial}
Figure \ref{fig:partial} shows two partial isomorphism types $\tau_1$ (left) and $\tau_2$ (right), 
where $R(\emph{ID}, A)$ is the only database relation and $\{x, y, z\}$ are 3 variables of type 
$R.\emph{ID}$. Solid lines are $=$-edges and dashed lines are $\neq$-edges.
In $\tau_1$, $(x, y)$ is connected with $=$ so
the edge $(x.A, y.A, =)$ is enforced by the key dependency. Missing edges between $(x.A, z.A)$ 
and $(y.A, z.A)$ indicate these connections are ``unknown''. 
$\tau_2$ is a full isomorphism type, which requires the graph to be complete so $(x.A, z.A)$ 
, $(y.A, z.A)$ and all pairs between $\{x, y, z\}$ and $\{x.A, y.A, z.A\}$ 
must be connected by either $=$ or $\neq$.
The $\neq$-edges between $\{x, y, z\}$ and $\{x.A, y.A, z.A\}$ are omitted in the figure for clarity.
\end{example}
\vspace*{-6mm}

We next define \emph{partial symbolic instances}. Intuitively, a partial symbolic instance consists of 
a partial isomorphism type capturing the connections of the current tuple of $\bar{x}$, as well as, for the
tuples present in the artifact relations, the represented isomorphism types $t$ and the count of tuples sharing $t$.

\vspace*{-2mm}
\begin{definition}
\label{def:psi}
A \textbf{partial symbolic instance} $I$ is a tuple $(\tau, \bar{c})$ where 
$\tau$ is a partial isomorphism type and $\bar{c}$ is a vector of 
$\mathbb{N}$ where each dimension of $\bar{c}$
corresponds to a unique partial isomorphism type.
\end{definition}
\vspace*{-2mm}
It turns out that most of the dimensions of $\bar{c}$ equal 0 in practice, so in implementation we only materialize a list of those dimensions with positive counter values.
We denote by $\apos(\bar{c})$ the set $\{\tau_S | \bar{c}(\tau_S) > 0\}$. 

Next, we define \emph{symbolic transitions} among partial symbolic instances by applying internal services.
First we need to define condition evaluation on partial isomorphism types.
Given a partial isomorphism type $\tau$, satisfaction of a condition $\phi$ in negation normal form\footnote{Negations are pushed down to leaf atoms.}
by $\tau$, denoted $\tau \models \phi$, is defined as follows:
\vspace*{-2mm}
\begin{itemize}\itemsep=0pt\parskip=0pt
\item $x \circ y$ holds in $\tau$ iff $(x, y, \circ) \in \tau$ for $\circ \in \{=, \neq\}$, 
\item for relation $R(\ID, A_1, \dots, A_m)$,
$R(x, y_1, \dots, y_m)$ holds in $\tau$ iff $(y_i, x.A_i, =) \in \tau$ for every $1 \leq i \leq m$,
\item $\neg R(x, y_1, \dots, y_m)$ holds in $\tau$ iff $(y_i, x.A_i, \neq) \in \tau$ for some $1 \leq i \leq m$, and
\item Boolean combinations of conditions are standard.
\end{itemize}
\vspace*{-2mm}

Notice that $\tau \not\models \phi$ might be due to missing edges in $\tau$ 
but not because of inconsistent edges, so it is possible to satisfy $\phi$ by filling in the missing edges.
This is captured by the notion of extension. We call $\tau'$ an \emph{extension} of $\tau$ if $\tau \subseteq \tau'$ and $\tau'$ is consistent, meaning that the edges in $\tau'$ do not imply any contradiction of (in)equalities. We denote by $\mathtt{eval}(\tau, \phi)$ the set of all \emph{minimal extensions} $\tau'$ of $\tau$
such that $\tau' \models \phi$. 
Intuitively, $\mathtt{eval}(\tau, \phi)$ contains partial isomorphism types
obtained by augmenting $\tau$ with a minimal set of constraints to satisfy $\phi$.

A symbolic transition is defined informally as follows (the full definition can be found in Appendix
\ref{app:semantics}).
To make a symbolic transition with a service $\sigma = (\pi, \psi, \bar{y}, \delta)$ 
from $I = (\tau, \bar{c})$ to $I' = (\tau', \bar{c}')$,
we first extend the partial isomorphism type $\tau$ to a new partial isomorphism type $\tau_0$ to satisfy the pre-condition $\pi$. 
Then the constraints on the propagated variables $\bar{y}$ are preserved 
by computing $\tau_1$, the projection of $\tau_0$ onto $\bar{y}$. 
Intuitively, the projection keeps only the expressions headed by variables in $\bar{y}$
and their connections. Finally, $\tau'$ is obtained by extending $\tau_1$
to satisfy the post-condition $\psi$. If $\delta$ is an insertion, then the counter that corresponds to the 
partial isomorphism type of the inserted tuple is incremented. 
If $\delta$ is a retrieval, then a partial isomorphism type $\tau_S$ with positive count is chosen 
nondeterministically and its count is decremented.
The new partial isomorphism type $\tau'$ is then extended with the constraints from $\tau_S$.
We denote by $\asucc(I)$ the set of possible successors of $I$ by taking one symbolic transition with any service $\sigma$.

\begin{figure}[!ht]
\centering
\includegraphics[width=0.50\textwidth]{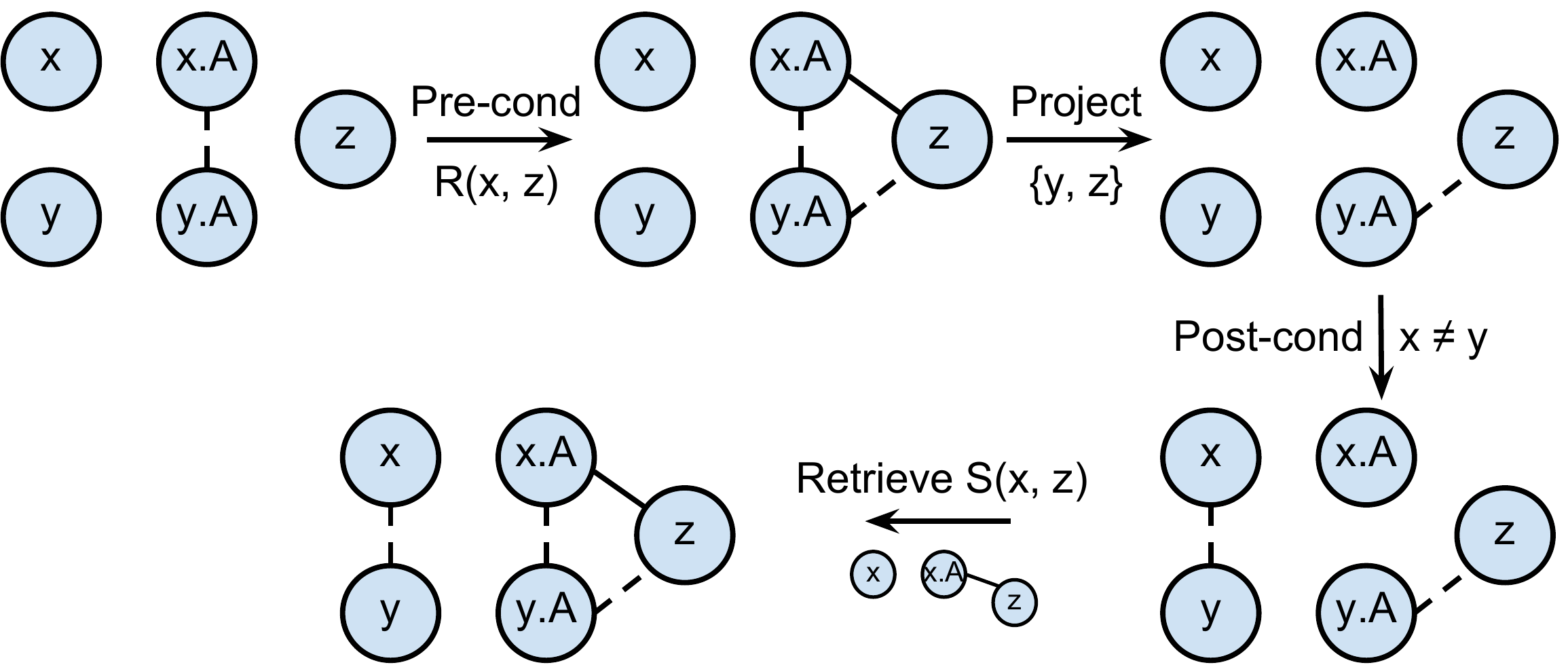}
\vspace*{-6mm}
\caption{Symbolic Transition}
\vspace*{-5mm}
\label{fig:transition}
\end{figure}

\vspace*{-1.5mm}
\begin{example}
Figure \ref{fig:transition} shows an example of symbolic transition. 
The DB schema is that of Example \ref{exm:partial}. 
The variables are $x, y$ of type $R.\emph{ID}$ and a non-ID variable $z$, with input variables $\{y, z\}$.
The applied service is $\sigma = (\pi : R(x, z), \psi : x \neq y, \bar{y} : \{y, z\}, \delta : \{-\cals(x, z)\})$.
First, the pre-condition $R(x, z)$ is evaluated so edge $(x.A, z, =)$ is added (top-middle).
Edge $(y.A, z, \neq)$ is also added so that the partial isomorphism type remains valid. 
Then variables $\{y, z\}$ are propagated, so the edges related to $x$ or $x.A$ are removed (top-right).
Next, we evaluate the post-condition $x \neq y$ so $(x, y, \neq)$ is added (bottom-right).
Finally, a tuple from $S$ is retrieved and overwrites $\{x, z\}$. Suppose the nondeterministically 
chosen $\tau_S$
contains a single edge $(x.A, z, =)$ (below the retrieve arrow). Then $\bar{c}(\tau_S)$ is decremented and $\tau_S$ is merged into
the final partial isomorphism type (bottom left). 
Note that if $\sigma$ contains an insertion of $+S(x, z)$ in $\delta$ instead of a retrieval, 
then the subgraph of $\tau_0$ (top-middle) projected to $\{x, z\}$ is inserted to $S$. 
The corresponding counter in $\bar{c}$ will be incremented by 1.
\end{example}
\vspace*{-2mm}

With symbolic transitions in place, verification works as follows. 
Informally, given a single-task HAS* $\Gamma$ and a LTL-FO property $\varphi$, one can check whether $\Gamma \models \varphi$
by constructing a new HAS* $\Gamma'$ obtained by combining $\Gamma$ with conditions in $\varphi$ and the \buchi\ 
automaton $B_{\neg \varphi}$ built from $\neg \varphi$. We can show that deciding whether $\Gamma \not\models \varphi$
reduces to checking whether an accepting state of $B_{\neg \varphi}$ is repeatedly reachable in $\Gamma'$. 
Verification therefore amounts to solving the following problem:
\vspace*{-1mm}
\begin{problem}{(Symbolic Repeated Reachability, or SRR)}
Given a HAS* $\Gamma$, an initial partial symbolic state $I_0$, and a condition $\phi$, 
is there a partial symbolic run $\{I_i\}_{0 \leq i \leq m < n}$ of $\Gamma$ such that 
$I_{i + 1} \in \asucc(I_i)$ for every $i \geq 0$, $\tau_m = \tau_n$, $\bar{c}_m \leq \bar{c}_n$ and $\tau_n \models \phi$?
\end{problem}

The condition $\phi$ above simply states that  $B_{\neg \varphi}$ is in one of its accepting states.

\subsection{The Classic Karp-Miller Algorithm} \label{sec:classic-km}

The SRR Problem 
defines an infinite search space due to the unbounded counter values,
so reduction to finite-state model checking is not possible.
Adapting the theory developed in~\cite{pods16} from symbolic representation based on
isomorphism types to symbolic representation based on partial isomorphism types, 
we can show that the symbolic transitions defined in Section~\ref{sec:partial}
can be modeled as a VASS whose states are the partial symbolic instances of 
Definition~\ref{def:psi}. Consequently, The SRR problem reduces to testing
(repeated) state reachability in this VASS. The benefit of the new approach is that this VASS is likely to have
much fewer states and counters than the one defined in~\cite{pods16}, because our search
materializes partial isomorphism types parsimoniously, by lazily expanding the current partial type $c$ using the (typically few)
constraints of the symbolic transition to obtain a successor $s$. The transition from $c$ to $s$ concisely represents
all the transitions from full-type expansions of $c$ to full-type expansions of $s$ (exponentially many in the number of ``unknown''
connections in $c$ and $s$), which in the worst case would be individually explored by the algorithm of~\cite{pods16}. 

The VERIFAS implementation of the (repeated) state reachability is based on a series of 
optimizations to the classic Karp-Miller algorithm \cite{karp-miller}. We describe the 
original Karp-Miller algorithm first, addressing the optimizations subsequently.

The Karp-Miller algorithm constructs a finite representation which over-approximates the
entire (potentially infinite) reachable VASS state space, called a  ``coverability set''
of states \cite{finkel1993minimal}. Any coverability set captures sufficient information about the global state space to
support global reasoning tasks, including repeated reachability.
In our context, the VASS states are the partial symbolic instances (PSIs) and 
a \emph{coverability} set is a finite set $\cali$ of PSIs, each reachable from the 
initial PSI $I_0$, 
such that for every reachable PSI $I = (\tau, \bar{c})$, 
there exists $I' = (\tau', \bar{c}') \in \cali$ with  $\tau = \tau'$ and 
$\bar{c} \leq \bar{c}'$. We say that $I'$ {\em covers} $I$, denoted $I \leq I'$. 
To represent counters that can increase forever, the coverability set also 
allows an extension of PSIs in which some of the counters can equal $\omega$. 
Recall that the ordinal $\omega$ is a special constant where $n < \omega$ for all $n \in \mathbb{N}$, $\omega \leq \omega$ 
and $\omega \pm 1 = \omega$. 

Since the coverability set $\cali$ is finite, we can effectively extract from it
the reachable $\tau_n$'s that satisfy  the condition $\phi$ (referring to the notation of
the SRR problem). To test whether $\tau_n$ is repeatedly reachable,
we can show that $I_n$ is repeatedly reachable iff $I_n$ is contained in a cycle consisting of only states in $\cali$ (proved in \cite{extended}). 
As a result, the repeatedly reachable $\tau_n$'s can be found by
constructing the transition graph among $\cali$ and computing its \emph{strongly connected components}.
A partial isomorphism type $\tau$ is repeatedly reachable if 
its corresponding PSI $I$ is included in a component containing a non-trivial cycle.

The Karp-Miller algorithm searches for a coverability set by materializing a finite part of
the (potentially infinite) VASS transition graph starting from the initial state
and executing transitions, pruning transitions to states that are covered by already
materialized states. The resulting transition subgraph is
traditionally called the \emph{Karp-Miller tree} (despite the fact that it is actually a DAG). 

In the notation of the SRR problem,
note that if at least one counter value strictly increases from
$I_m$ to $I_n$ ($\bar c_m^i < \bar c_n^i$ for some dimension $i$),
then the sequence of transitions from $m$ to $n$
can repeat indefinitely, periodically reaching states with the same 
partial isomorphism type $\tau_n$, but with ever-increasing associated counter values in
dimension $i$ (there are infinitely many such states).
In the limit, the counter value becomes $\omega$ so a coverability set must include
a state $(\tau_n, \bar c)$ with $\bar c^i = \omega$, covering these infinitely
many states.

Finite construction of the tree is possible due to a special \emph{accelerate} operation that skips
directly to a state with $\omega$-valued counters, avoiding the construction of
the infinitely many states it covers.
Adapted to our context, 
when the algorithm detects a path $\{I_i\}_{0 \leq i \leq m < n}$ in the tree 
where $I_m \leq I_n$, the accelerate operation replaces in $I_n$ the values 
of $\bar{c}_n(\tau_S)$ with $\omega$ for every $\tau_S$ where 
$\bar{c}_m(\tau_S) < \bar{c}_n(\tau_S)$.

We outline the details in Algorithm \ref{alg:karp-miller}, which outputs the Karp-Miller
tree $\calt$.
We denote by $\mathtt{ancestors}(I)$ the set of ancestors of $I$ in $\calt$. 
Given a set $\cali$ of states and a state $I' = (\tau', \bar{c}')$,
the accelerate function is defined as $\mathtt{accel}(\cali, I') = (\tau', \bar{c}'')$ where for every $\tau_S$,
$\bar{c}''(\tau_S) = \omega$ if there exists $(\tau, \bar{c}) \in \cali$ such that 
$\tau = \tau'$, $\bar{c} \leq \bar{c}'$ and $\bar{c}(\tau_S) < \bar{c}'(\tau_S)$. 
Otherwise, $\bar{c}''(\tau_S) = \bar{c}'(\tau_S)$.

\vspace*{-3mm}
\begin{algorithm}[!ht]
\SetKwInOut{Input}{input}
\SetKwInOut{Output}{output}
\SetKwInOut{Variables}{variables}
\Input{Initial instance $I_0$}
\Output{$\calt$, the Karp-Miller tree}
\Variables{$W$, set of states waiting to be explored}
$W \leftarrow \{I_0\}$, $\calt \leftarrow (\{I_0\}, \emptyset)$\;
\While{$W \neq \emptyset$} {
	Remove a state $I$ from $W$\;
	\For{$I' \in \asucc(I)$} {
		$I'' \leftarrow \mathtt{accel}(\mathtt{ancestors}(I), I')$\;
		\If{$I'' \not\in \calt \lor I'' \in W$}{
			Add edge $(I, I'')$ to $\calt$\;
			$W \leftarrow W \cup \{I''\}$\;
		}
	}
}
Return $\calt$\;
\caption{Karp-Miller Tree Search Algorithm}
\label{alg:karp-miller}
\end{algorithm}

\subsection{Optimization with Monotone Pruning}\label{sec:pruning}

The original Karp-Miller algorithm is well-known to be inefficient in practice due to state explosion.
To improve performance, various techniques have been proposed.
The main technique we adopted in VERIFAS is based on pruning the Karp-Miller tree
by monotonicity. Intuitively, when a new state $I = (\tau, \bar{c})$ is generated during the search,
if there exists a visited state $I'$ where $I \leq I'$,
then $I$ can be discarded because for every state $\tilde{I}$ reachable from $I$,
there exists a reachable state $\tilde{I}'$ starting from $I'$ such that $\tilde{I} \leq \tilde{I}'$ 
by applying the same sequence of transitions that leads $I$ to $\tilde{I}$.
For the same reason, if $I \geq I'$, then $I'$ and its descendants can be pruned from the tree. 
However, correctness of pruning is sensitive to the order of application of these rules
(for example, as illustrated in \cite{karp-miller-pruning-more}, 
application of the rules in a breadth-first order may lead to incompleteness). 
The problem of how to apply the rules without losing completeness was studied in \cite{karp-miller-pruning-more, karp-miller-pruning}
and we adopt the approach in \cite{karp-miller-pruning}. 
More specifically, Algorithm \ref{alg:karp-miller} is extended by keeping track of a set 
$\mathtt{act}$ of ``active'' states and adding the following changes:

\vspace*{-1mm}
\begin{itemize}\itemsep=0pt\parskip=0pt
\item Initialize $\mathtt{act}$ with $\{I_0\}$;
\item In line 3, choose the state from $W \cap \mathtt{act}$;
\item In line 5, $\mathtt{accel}$ is applied on $\mathtt{ancestors}(I) \cap \mathtt{act}$;
\item In line 8, $I''$ is not added to $W$ if there exists $\hat{I} \in \mathtt{act}$ such that $I'' \leq \hat{I}$;
\item When $I''$ is added to $W$, remove from $\mathtt{act}$ every state $\hat{I}$ and its descendants 
for $\hat{I} \leq I''$ and $\hat{I}$ is either active or not an ancestor of $I''$. Add $I''$ to $\mathtt{act}$.
\end{itemize}
\vspace*{-1mm}

\subsection{A Novel, More Aggressive Pruning} \label{sec:aggressive-pruning}
We generalize the comparison relation $\leq$ 
of partial symbolic instances to achieve more aggressive pruning of the 
explored transitions.
The novel comparison is based on the insight that a state $I$ can be pruned in favor of 
$I'$ as long as every partial isomorphism type reachable from $I$ is also reachable from 
$I'$. 
So $I = (\tau, \bar{c})$ can be pruned by $I' = (\tau', \bar{c}')$ if $\tau'$ 
is ``less restrictive'' than $\tau$ (or $\tau$ implies $\tau'$), and 
for every occurrence of $\tau_S$ in $\bar{c}$, there exists a corresponding occurrence of 
$\tau_S'$ in $\bar{c}'$ such that $\tau_S'$ is ``less restrictive''
than $\tau_S$. Formally, given partial isomorphism types $\tau$ and $\tau'$, 
$\tau$ implies $\tau'$, denoted as $\tau \models \tau'$, iff $\tau' \subseteq \tau$.
We replace the coverage relation $\leq$ on partial symbolic instances with 
a new binary relation $\preceq$ as follows. 
\vspace*{-1mm}
\begin{definition}\label{def:preceq}
Given two partial symbolic states $I = (\tau, \bar{c})$ and $I' = (\tau', \bar{c}')$,
$I \preceq I'$ iff $\tau \models \tau'$ and there exists 
$f: \apos(\bar{c}) \times \apos(\bar{c}') \mapsto \mathbb{N} \cup \{\omega\} $ such that
\vspace*{-1mm}
\begin{itemize}\itemsep=0pt\parskip=0pt
\item $f(\tau_S, \tau_S') > 0$ only when $\tau_S \models \tau_S'$,
\item for every $\tau_S$, $\sum_{\tau_S'} f(\tau_S, \tau_S') = \bar{c}(\tau_S)$ and
\item for every $\tau_S'$, $\sum_{\tau_S} f(\tau_S, \tau_S') \leq \bar{c}'(\tau_S')$.
\end{itemize}
\end{definition}
\vspace*{-2mm}

Intuitively, $f$ describes a one-to-one mapping from tuples stored in 
the artifact relations in $I$ to tuples in $I'$. 
$f(\tau_S, \tau_S') = k$
means that there are $k$ tuples in $I$ of partial isomorphism type $\tau_S$ that are mapped to 
$k$ tuples in $I'$ of type $\tau_S'$. The condition $\tau_S \models \tau_S'$
guarantees that each tuple in $I$ is mapped to one  in $I'$ of a less restrictive type.

\begin{figure}[!ht]
\centering
\includegraphics[width=0.48\textwidth]{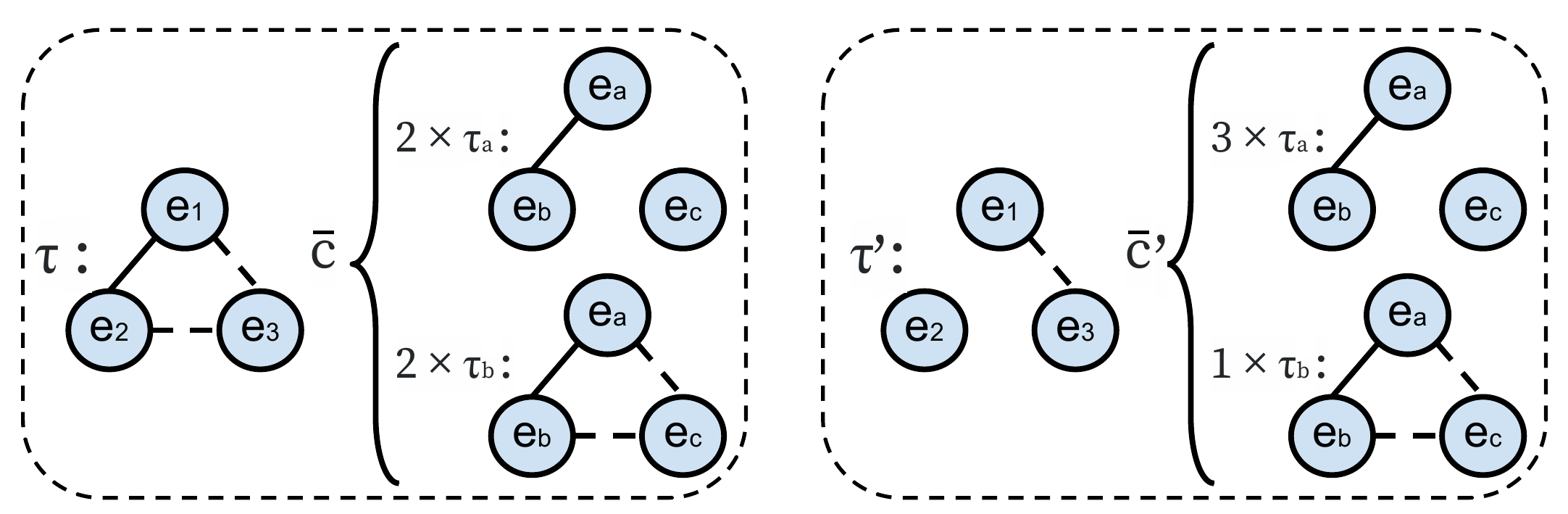}
\vspace*{-6mm}
\caption{Illustration of $\preceq$}
\vspace*{-3mm}
\label{fig:pruning}
\end{figure}

\begin{example}
Consider the two PSIs $I = (\tau, \bar{c} = \{\tau_a : 2, \tau_b : 2\})$ (left) and 
$I' = (\tau', \bar{c}' = \{\tau_a : 3, \tau_b : 1\})$ (right) shown in Figure \ref{fig:pruning}. 
Since $\tau \neq \tau'$ and $\bar{c}(\tau_b) > \bar{c}'(\tau_b)$, $I \leq I'$ does not hold. 
However, any sequence of symbolic transitions applicable
starting from $I$ can also be applied starting from $I'$, because if the conditions
imposed by these transitions do not conflict with those in $I$,  
then they won't conflict with the subset thereof in $I'$. Consequently, $I$ can be pruned 
if $I'$ is found during the search. This fact is detected by $\preceq$: 
$I \preceq I'$ holds since  $\tau \models \tau'$ and we can construct $f$ as 
$f(\tau_a, \tau_a) = 2$ and $f(\tau_b, \tau_b) = f(\tau_b, \tau_a) = 1$
since $\tau_b \models \tau_a$.  
\end{example}
\vspace*{-3mm}

Note that one can efficiently test whether $I \preceq I'$ by reduction to the Max-Flow problem
over a flow graph $F$ with node set $\apos(\bar{c}) \cup \apos(\bar{c}') \cup \{s,t\}$, 
with $s$ a source node and $t$ a sink node. 
For every $\tau_S \in \apos(\bar{c})$, there is an edge
from $s$ to $\tau_S$ with capacity $\bar{c}(\tau_S)$. 
For every $\tau_S' \in \apos(\bar{c}')$,
there is an edge from $\tau_S'$ to $t$ with capacity $\bar{c}'(\tau_S)$.
For every pair of $\tau_S, \tau_S'$, there is an edge from $\tau_S$ to $\tau_S'$ 
with capacity $\infty$ if $\tau_S' \models \tau_S$.
We can show that $F$ has a max-flow equal to $\sum_{\tau_S} \bar{c}(\tau_S)$ 
if and only if $I \preceq I'$.

The same idea can also be applied to the $\mathtt{accel}$ function. 
Formally, given $\cali$ and $I' = (\tau', \bar{c}')$, the new accelerate function \\
$\mathtt{accel}(\cali, I') = (\tau', \bar{c}'')$ where $\bar{c}''(\tau_S') = \omega$ if
there exists $I \in \cali$ such that $I \preceq I'$ and there exists mapping  
$f$ satisfying the conditions in Definition \ref{def:preceq} and 
$\sum_{\tau_S} f(\tau_S, \tau_S') < \bar{c}'(\tau_S')$. Otherwise $\bar{c}''(\tau_S') = \bar{c}'(\tau_S')$.


\subsection{Data Structure Support} \label{sec:data-structure}

The above optimization relies on two important operations applied every time a new state is explored:
given the set of active states $\mathtt{act}$ and a partial symbolic state $I$,
(1) compute the set $\{ I' | I' \preceq I \land I' \in \mathtt{act}\}$ and (2) check whether
there exists $I' \in \mathtt{act}$ such that $I \preceq I'$.
As each test of $\preceq$ might require an expensive operation of computing the max-flow,
when $|\mathtt{act}|$ is large, checking whether $I' \preceq I$ (or $I \preceq I'$) for every $I' \in \mathtt{act}$
would be too time-consuming. 

We start with the simple case where $\bar{c} = \bar{0}$ in all $I$'s. Then to test whether $I \preceq I'$ 
for $I = (\tau, \bar{c})$ and $I' = (\tau', \bar{c}')$ is to test whether
$\tau' \subseteq \tau$. When the partial isomorphism types are stored as sets of edges,
we can accelerate the two operations with data structures that 
support fast subset (superset) queries: given a collection $\calc$ of sets and a query set $q$,
find all sets in $\calc$ that are subsets (supersets) of $q$.
The standard solutions are to use Tries for superset queries \cite{rivest1976partial} 
and Inverted Lists for subset queries \cite{manning2008introduction}.

The same idea can be applied to the general case where $\bar{c} \geq 0$
to obtain over-approximations of the precise results. 
Given $I = (\tau, \bar{c})$, we let $E(I)$ be the set of edges in $\tau$
or any $\tau_S$ where $\bar{c}(\tau_S) > 0$.
Then given $I$ and $I'$, $I \preceq I'$ implies $E(I') \subseteq E(I)$. 
We build the Trie and Inverted Lists indices such that for a given query $I$, 
they return a candidate set $\cali^{\preceq}$ and a candidate set $\cali^{\succeq}$ 
where $\cali^{\preceq}$ contains all $I'$ from $\mathtt{act}$ such that $E(I') \subseteq E(I)$ and
$\cali^{\succeq}$ contains all $I'$ such that $E(I) \subseteq E(I')$.
Then it suffices to test each member in the candidate sets for $I \preceq I'$ and $I \succeq I'$
to obtain the precise results of operations (1) and (2).


%

\subsection{Optimization with Static Analysis} \label{sec:static}

Next, we introduce our optimization based on static analysis.
At a high level, we notice that in real workflow examples, 
some constraints in conditions of the specification and the property are irrelevant to the result of verification
because they can never cause violations when conditions are evaluated in a symbolic run.
Such conditions can be ignored to reduce the number of symbolic states.
For example, for a constraint $x = y$ in the specification,
if $x \neq y$ does not appear anywhere else and cannot be implied
by other constraints, then $x = y$ can be safely removed from any partial isomorphism types
without affecting the result of the verification algorithm.
Our goal is to detect all such constraints by statically analyzing the HAS* and the LTL-FO property.
Specifically, we analyze the constraint graph consisting of all possible 
``$=$'' and ``$\neq$'' constraints that can potentially be added to any 
partial isomorphism types in symbolic transitions of the HAS* $\Gamma$ or when checking condition $\phi$ 
(refer to the notation in the SRR problem). 

\vspace*{-1mm}
\begin{definition} \label{def:eqgraph}
The constraint graph $G$ of $(\Gamma, \phi)$ is a labeled undirected graph over 
the set of all expressions $\cale$ with the following edges. 
For every atom $a$ that appears in a condition of $\Gamma$ or $\phi$ 
in negation normal form, if $a$ is
\vspace*{-1.5mm}
\begin{itemize}\itemsep=0pt\parskip=0pt
\item $(x = y)$, then $G$ contains $(x.w, y.w, =)$ for all sequences $w$ where $\{x.w, y.w\} \subseteq \cale$,
\item $(x \neq y)$, then $G$ contains $(x, y, \neq)$,
\item $R(x, y_1, \dots, y_m)$, then $G$ contains $(x.A_i.w, y_i.w, =)$ for all $i$ and sequences $w$ where 
$\{x.A_i.w, y_i.w\} \subseteq \cale$, and
\item $\neg R(x, y_1, \dots, y_m)$, then $G$ contains $(x.A_i, y_i, \neq)$ for all $i$.
\end{itemize}
\vspace*{-1.5mm}
For any subgraph $G'$ of $G$, $G'$ is {\em consistent} if 
the edges in $G'$ do not imply any contradiction, meaning that there is no path of $=$-edges
connecting two distinct constants or two expressions connected by an $\neq$-edge.

An edge $e$ of $G$  is {\em non-violating} if for every consistent subgraph $G'$, 
$G' \cup \{e\}$ is also consistent.
\end{definition}
\vspace*{-1.5mm}

Intuitively, by collecting the edges described above, 
the constraint graph $G$ becomes an over-approximation of the reachable partial isomorphism types.
Thus any edge in $G$ that is non-violating is also non-violating in any
reachable partial isomorphism type.
So our goal is to find all the non-violating edges in $G$, 
since they can be ignored in partial isomorphism types to reduce the size of the search space.

Non-violating edges can be identified efficiently in polynomial time. 
Specifically, an edge $(u, v, \neq)$ is non-violating if
$u$ and $v$ belong to different connected components of $=$-edges of $G$.
An $=$-edge $e$ is non-violating if there is no path $u \goto{ } v$ of $=$-edges containing $e$
for any $(u, v, \neq) \in G$ or $(u,v)$ being two distinct constants. 
This can be checked efficiently by computing the biconnected components of the $=$-edges~\cite{tarjan1972depth}. 
We omit the details here.

\begin{figure}[!ht]
\centering
\includegraphics[width=0.48\textwidth]{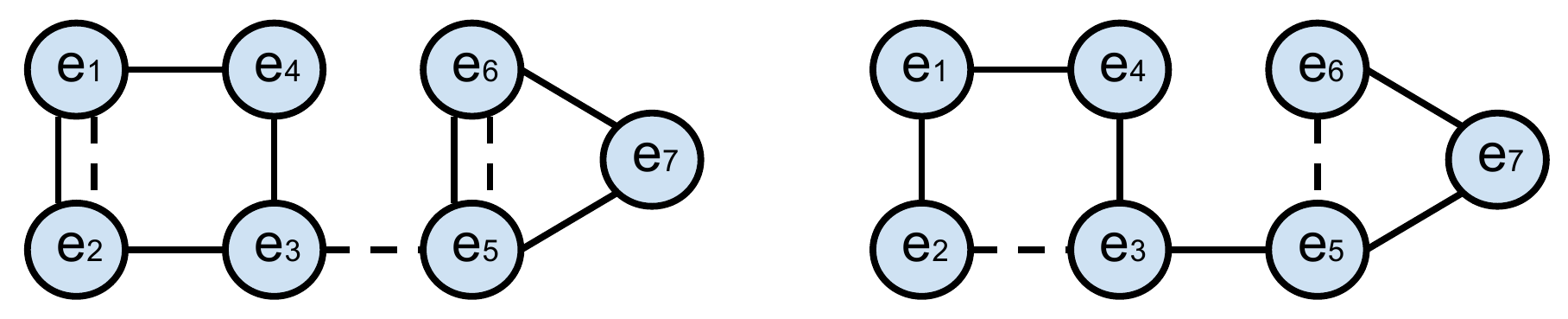}
\vspace*{-4mm}
\caption{Non-violating Edges}
\vspace*{-4mm}
\label{fig:nonviolating}
\end{figure}

\vspace*{-2mm}
\begin{example}
Consider the two constraint graphs $G_1$ and $G_2$ in Figure \ref{fig:nonviolating}. 
In $G_1$ (left), $(e_3, e_5)$ is a non-violating $\neq$-edge because 
$e_3$ and $e_5$ belong to two different connected components of $=$-edges 
($\{e_1, e_2, e_3, e_4\}$ and $\{e_5, e_6, e_7\}$ respectively).
In $G_2$ (right), $(e_3, e_5)$ is a non-violating $=$-edge because 
$(e_3, e_5)$ is not on any simple path of $=$-edges connecting 
the two ends of any $\neq$-edges (i.e. $(e_2, e_3)$ and $(e_5, e_6)$).
\end{example}
\vspace*{-6mm}

\subsection{Extension to Repeated-Reachability} \label{sec:repeated}

Recall from Section~\ref{sec:partial} 
that providing full support for verifying LTL-FO properties requires solving the
repeated state reachability problem.
It is well-known that for VASS, the coverability set $\cali$ extracted from the tree $\calt$
constructed by the classic Karp-Miller algorithm can be used to 
identify the repeatedly reachable partial isomorphism types (see Section~\ref{sec:classic-km}).
The same idea can be extended to the Karp-Miller algorithm with monotone pruning (Section \ref{sec:pruning}),
since the algorithm is guaranteed to construct a coverability set.

However, the Karp-Miller algorithm equipped with our $\preceq$-based pruning
(Section~\ref{sec:aggressive-pruning}) explores fewer states due to the more aggressive pruning, and
it turns out that the resulting coverability set $\cali^\preceq$ is incomplete to determine whether a state is repeatedly reachable.
We can no longer guarantee that a repeatedly reachable state is only contained in a cycle of states in $\cali^\preceq$. 
We can nevertheless show that the completeness of the search for repeatedly reachable states
can be restored by developing our own extraction technique which compensates for the
overly aggressive $\preceq$-based pruning. The technical development is subtle and relegated to
Appendix \ref{app:karp-miller}.
As confirmed by our experimental results, the additional overhead is acceptable.

%% file: experiment.tex
\section{Experimental Evaluation} \label{sec:experiment}

We evaluated the performance of VERIFAS using both real-world and synthetic artifact specifications.

\subsection{Setup and Benchmark}

\noindent
\textbf{The Real Set}
We built an artifact system benchmark by rewriting in HAS* a sample of real-world BPMN workflows published 
at the official BPMN website \cite{bpmn}, which provides 36 workflows of non-trivial size. 
To rewrite these workflows in HAS*, we manually added the database schema, 
artifact variables/relations, and services for updating the data.
HAS* is sufficiently expressive to specify 32 of the 36 BPMN workflows.
The remaining 4 cannot be expressed in HAS* because they involve computing aggregate 
functions or updating unboundedly many tuples of the artifact relations, which is not supported 
in the current model. We will consider such features in our future work.

\vspace*{1mm}
\noindent
\textbf{The Synthetic Set }
Since we wished to stress-test VERIFAS,  we also randomly generated a set of HAS*
specifications of increasing complexity. 
All components of each specification, including DB schema, task hierarchy, variables, services and 
conditions, were generated fully at random for a certain size. 
We provide in Appendix \ref{app:experiment} more details on how each specification is generated.
Those with empty state space due to unsatisfiable conditions were removed from the benchmark.
Table \ref{tab:stat} shows some statistics of the two sets of specifications.
(\#Relations, \#Tasks, etc. are averages over the real / synthetic sets of workflows.) 

\vspace*{-2mm}
\setlength{\tabcolsep}{3pt}
\begin{table}[!ht]
\centering
  \begin{tabular}{cccccc}
    \toprule
Dataset   & Size   & \#Relations & \#Tasks & \#Variables & \#Services \\ 
      \midrule
Real      & 32     & 3.563   & 3.219   & 20.63   & 11.59 \\ 
Synthetic & 120    & 5       & 5       & 75      & 75 \\ 
    \bottomrule
  \end{tabular}
\vspace*{-2mm}
\caption{Statistics of the Two Sets of Workflows}
\label{tab:stat}
\end{table}
\vspace*{-5mm}

\vspace*{1mm}
\noindent
\textbf{LTL-FO Properties } On each workflow of both sets,
we run our verifier on a collection of 12 LTL-FO properties of the root task constructed 
using templates of real propositional LTL properties.
The LTL properties are all the 11 examples of safety, liveness and fairness properties collected from 
a standard reference paper \cite{sistla1994safety} and an additional property $\mathtt{False}$
used as a baseline when comparing the performance of VERIFAS on different classes of LTL-FO properties.
We list all the templates of LTL properties in Table \ref{tab:ltlfo}.
To see why we choose $\mathtt{False}$ as the baseline property,
recall from Section \ref{sec:karp-miller} that the verifier's running time is mainly determined by the
size of the reachable symbolic state space (VERIFAS first computes all reachable symbolic states
--represented by the coverability set-- then identifies the repeatedly-reachable ones).
The reachable symbolic state space can be conceptualized as the cross-product between the reachable symbolic
state space of the HAS* specification (absent any property) and the \buchi\ automaton of the property.
When the LTL-FO property is $\mathtt{False}$, 
the generated \buchi \ automaton is of the simplest form (a single accepting state within a loop), so
it has no impact on the cross-product size, unlike more complex properties.

In each workflow, we generate an LTL-FO property for each template by 
replacing the propositions with 
FO conditions chosen from the pre-and-post conditions and their sub-formulas.
Note that by doing so, the generated LTL-FO properties on the real workflows
are combinations of real propositional LTL properties and real FO conditions, and so 
are close to real-world LTL-FO properties.

\vspace*{1mm}
\noindent
\textbf{Baseline} We compare VERIFAS with a simpler implementation built on top of
Spin, a widely used software verification tool \cite{spin}. 
Building such a verifier is a challenging task in its own right since Spin is essentially 
a finite-state model checking tool and hence is 
incapable of handling data of unbounded size, present in the HAS* model. 
We managed to build a Spin-based verifier 
supporting a restricted version of our model that does not handle updatable artifact relations. 
As the read-only database can still have unbounded size and domain, 
the Spin-based implementation requires nontrivial translations and optimizations, 
which are presented in detail in \cite{spinarxive}.

\vspace*{1mm}
\noindent
\textbf{Platform}
We implemented both verifiers in \texttt{C++} with Spin version 6.4.6
for the Spin-based verifier.
All experiments were performed on a Linux server with a quad-core Intel i7-2600 CPU and 16G memory.
For each specification, we ran our verifiers to test each of the 12 generated LTL-FO properties,
resulting in 384 runs for the real set and 1440 runs for the synthetic set.
Towards fair comparison, since the Spin-based verifier (Spin-Opt) cannot handle artifact relations, 
in addition to running our full verifier (VERIFAS), 
we also ran it with artifact relations ignored (VERIFAS-NoSet).
The timeout limit of each run was set to 10 minutes and the memory limit was set to 8G. 

\subsection{Experimental Results}

\noindent
\textbf{Performance }
Table \ref{tab:performance} shows the results on both sets of workflows. 
The Spin-based verifier achieves acceptable performance in the real set, with
an average elapsed time of a few seconds and only 3 timeouts.
However, it failed in a large number of runs (440/1440) in the stress-test using synthetic specifications.
On the other hand, both VERIFAS and VERIFAS-NoSet achieve 
average running times within 0.3 second and no timeout on the real set,
and the average running time is within seconds on the synthetic set, with only 19 timeouts over 1440 runs.
The presence of artifact relations introduced an acceptable amount of performance overhead, 
which was negligible in the real set and less than 60\% in the synthetic set.
Compared with the Spin-based verifier, VERIFAS is $>$10x faster in average running time
and scales significantly better with workflow complexity.

The timeout runs on the synthetic workflows are all due to state explosion with a
state space of size $\sim$3$\times10^4$. The reason is that though unlikely in practice,
it is still possible that the reached partial isomorphism types can degenerate to
full isomorphism types, and in this case our state-pruning optimization does not reduce 
the number of reached states.

\setlength{\tabcolsep}{3pt}
\begin{table}
  \centering
  \begin{tabular}{ccccc}
    \toprule
    \multirow{2}{*}{Verifier} &
      \multicolumn{2}{c}{Real} &
      \multicolumn{2}{c}{Synthetic} \\
      & {Avg(Time)} & {\#Fail}& {Avg(Time)} & {\#Fail} \\
      \midrule
    Spin-Opt & 2.97s          & 3  & 83.983s          & 440  \\
    VERIFAS-NoSet & \textbf{.229s} & 0  & \textbf{6.983s} & 19  \\
    VERIFAS       & \textbf{.245s} & 0  & \textbf{11.01s} & 16  \\
    \bottomrule
  \end{tabular}
\caption{Average Elapsed Time and Number of Failed Runs (\#Fail) due to Timeout or Memory Overflow}
\vspace{-6mm}
\label{tab:performance}
\end{table}

\vspace*{1mm}
\noindent
\textbf{Cyclomatic Complexity }
To better understand the scalability of VERIFAS, we also measured verification time
as a function of workflow complexity, adopting a metric called
{\em cyclomatic complexity}, which is widely used in measuring
complexity of program modules \cite{cyclomatic2}. 
For a program $P$ with control-flow graph $G(V, E)$, 
the cyclomatic complexity of $P$ equals $|E| - |V| + 2$. 
We adapt this measure to HAS* specifications as follows. 
Given a HAS* specification $\cala$, a control flow graph of $\cala$ can be obtained by selecting a task $T$ of $\cala$ 
and a non-id variable $x \in \bar{x}^T$ and projecting all services of $T$ onto $\{x\}$. 
The resulting services contain only $x$ and constants and thus can be viewed
as a transition graph with $x$ as the state variable. The cyclomatic complexity of $\cala$, 
denoted as $M(\cala)$, is defined as the maximum cyclomatic complexity 
over all the possible control-flow graphs of $\cala$ (corresponding to all possible projections).

\begin{figure}[!t]
\centering
\includegraphics[width=0.46\textwidth]{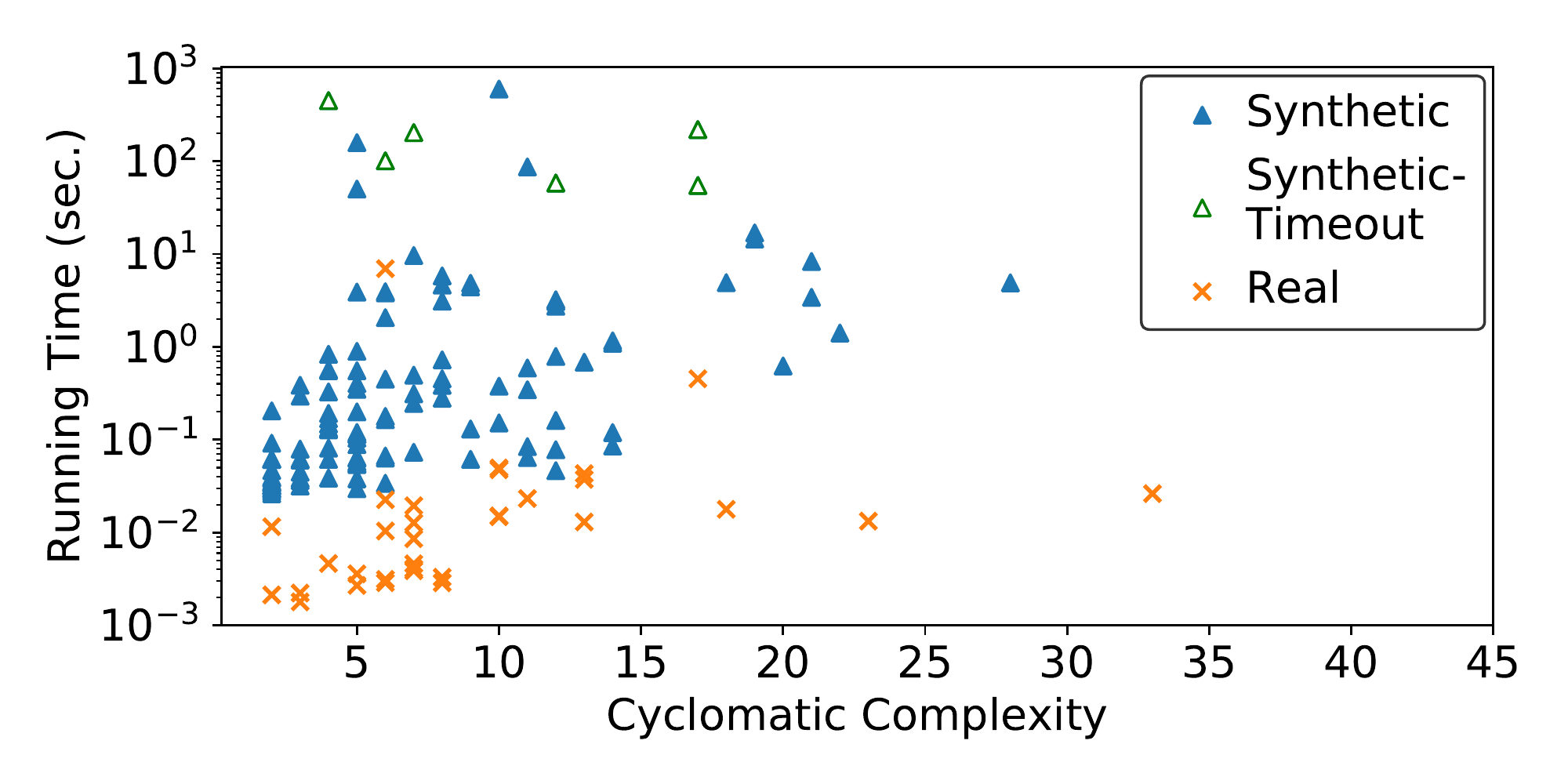}
\vspace*{-5mm}
\caption{Average Running Time vs. Cyclomatic Complexity}
\vspace*{-6mm}
\label{fig:cyc}
\end{figure}

Figure \ref{fig:cyc} shows that the verification time increases exponentially with the cyclomatic complexity,
thus confirming the pertinence of the measure to predicting verification complexity,
where the verification time of a workflow is measured by the average running time 
over all the runs of its LTL-FO properties.
According to \cite{cyclomatic2}'s recommendation, for a program to remain readable and testable, its 
cyclomatic complexity should not exceed 15.
Among all the 138 workflows with cyclomatic complexity at most 15, VERIFAS successfully
verified 130/138 ($\sim$94\%) of them within 10s and only 4 instances have timeout runs (marked as hollow triangles in Figure \ref{fig:cyc}).
For specifications with complexity above 15, only 2/14 instances have timeout runs.

%
Typically, for the same cyclomatic complexity, the real workflows can be verified
faster compared to the synthetic workflows. This is because the search space of 
the synthetic workflows is likely to be larger because there are more variables and transitions.

%

\vspace*{1mm}
\noindent
\textbf{Impact of Optimizations}
We studied the effect of our optimization techniques:
state pruning (SP, Section~\ref{sec:aggressive-pruning}),  
data structure support (DSS, Section~\ref{sec:data-structure}),
and static analysis (SA, Section~\ref{sec:static}).
For each technique, we reran the experiment with the optimization turned off and measured the speedup
by comparing the elapsed verification time with the original elapsed time. 
Table \ref{tab:speedup} shows the average speedups of each optimization on both datasets.
Since some instances have extreme speedups (over 10,000x), simply averaging could be misleading,
so we also present the trimmed averages of the speedups (i.e. removing the top/bottom 5\% speedups before averaging) 
to exclude the extreme values.

Table \ref{tab:speedup} shows that the effect of state pruning is the most significant in both sets of workflows, 
with an average (trimmed) speedup of $\sim$25x and $\sim$127x in the real and synthetic set respectively.
The static analysis optimization is more effective in the real set (1.4x improvement) 
but its effect in the synthetic set is less pronounced. It creates a small amount (7\%) of overhead in most cases,
but significantly improves the running time of a single instance, resulting in the huge gap between the normal average
speedup and the trimmed average speedup. 
The explanation to this phenomenon is that the workflows in the real set are more ``sparse" in general, 
which means there are fewer comparisons within a subset of variables so a larger number of 
comparisons can be pruned by static analysis. 
Finally, the data-structure support provides $\sim$1.2x and $\sim$1.6x average speedup in each set respectively.
Not surprisingly, the optimization becomes more effective as the size of the state space increases.

\vspace*{-1mm}
\setlength{\tabcolsep}{4pt}
\begin{table}[!ht]
\centering
  \begin{tabular}{ccccccc}
    \toprule
    \multirow{2}{*}{Dataset} &
      \multicolumn{2}{c}{SP} &
      \multicolumn{2}{c}{SA} &
      \multicolumn{2}{c}{DSS} \\
      & {Mean} & {Trim.}& {Mean} & {Trim.} & {Mean} & {Trim.} \\
      \midrule
Real      & 1586.54x & {24.69x} & 1.80x & {1.41x} & 1.87x & {1.24x} \\ 
Synthetic & 322.03x & {127.35x} & {28.78x} & 0.93x & 2.72x & {1.58x} \\ 
    \bottomrule
  \end{tabular}
\vspace{-2mm}
\caption{Mean and Trimmed Mean (5\%) of Speedups}
\label{tab:speedup}
\vspace{-6mm}
\end{table}


\vspace*{2mm}
\noindent
\textbf{Overhead of Repeated-Reachability } We evaluated the overhead of computing the
set of repeatedly-reachable states from the coverability set (Section \ref{sec:repeated})
by repeating the experiment with the repeated reachability module turned off.
Compared with the turned off version, the full verifier has an average overhead of \textbf{19.03\%} on the real set
and \textbf{13.55\%} overhead on the synthetic set (overheads are computed over the non-timed-out runs).


\vspace*{1mm}
\noindent
\textbf{Effect of Different Classes of LTL-FO Properties }
Finally, we evaluate how the structure of LTL-FO properties affects the performance of VERIFAS.
Table \ref{tab:ltlfo} lists all the LTL templates used in generating the LTL-FO properties
and their intuitive meaning stated in \cite{sistla1994safety}.
For each template and for each set of workflows, 
we measure the average running time over all the runs with LTL-FO properties generated using the template.
Table \ref{tab:ltlfo} shows that
for each class of properties, the average running time is within 2x of the
average running time for the simplest non-trivial property $\mathtt{False}$.
This is much better than the theoretical upper bound, 
which is linear in size of the \buchi \ automaton of the LTL formula.
Some properties even have a shorter running time because, 
although the space of partial symbolic instances is enlarged by the \buchi \ automaton, 
more of the states may become unreachable due to the additional
constrains imposed by the LTL-FO property.

\vspace*{-1mm}
\begin{table}[!ht]
\centering
  \begin{tabular}{cccc}
    \toprule
      Templates for LTL-FO  & Meaning & Real & Synthetic \\
      \midrule
$\mathtt{False}$ & Baseline & 0.26s & 10.13s \\
$\mathbf{G} \varphi$ & Safety & 0.28s & 10.26s \\
$(\neg \varphi \ \mathbf{U} \ \psi )$ & Safety & 0.28s & 16.13s \\
$(\neg \varphi \mathbf{U} \psi ) \land \mathbf{G} (\varphi \rightarrow \mathbf{X}(\neg \varphi \mathbf{U} \psi ) )$ & Safety & 0.30s & 10.79s \\
$\mathbf{G} (\varphi \rightarrow ( \psi \lor \mathbf{X} \psi \lor \mathbf{X} \mathbf{X} \psi) )$ & Safety & 0.29s & 12.07s \\
$\mathbf{G}(\varphi \lor \mathbf{G}(\neg \varphi))$ & Safety & 0.30s & 12.17s \\
$\mathbf{G} ( \varphi \rightarrow \mathbf{F} \psi )$ & Liveness & 0.29s & 16.81s \\
$\mathbf{F} \varphi $ & Liveness & 0.02s & 6.44s \\
$\mathbf{GF} \varphi \rightarrow \mathbf{GF} \psi$ & Fairness & 0.30s & 14.09s \\
$\mathbf{GF} \varphi $ & Fairness & 0.28s & 6.91s \\
$\mathbf{G} (\varphi \lor \mathbf{G} \psi)$ & Fairness & 0.05s & 9.64s \\
$\mathbf{FG} \varphi \rightarrow \mathbf{GF} \psi$ & Fairness & 0.28s & 6.75s \\
    \bottomrule
  \end{tabular}
\caption{Average Running Time of Verifying Different Classes of LTL-FO Properties
}
\label{tab:ltlfo}
\vspace{-6mm}
\end{table}

%% file: related-work.tex
\section{Related Work}\label{sec:related}

The artifact verification problem has previously been studied mainly from a theoretical perspective.
As mentioned in Section \ref{sec:intro}, fully automatic artifact verification is a challenging problem
due to the presence of unbounded data. To deal with the resulting infinite-state system, 
we developed in \cite{DHPV:ICDT:09} a symbolic approach allowing a reduction to finite-state model checking and
yielding a {\sc pspace} verification algorithm for the simplest variant of the
model (no database dependencies and uninterpreted data domain).  
In \cite{tods12} we extended our approach to allow for database dependencies and numeric data testable by arithmetic constraints. 
The symbolic approach developed in \cite{DHPV:ICDT:09} and its extension to HAS \cite{pods16}
provides the theoretical foundation for VERIFAS.

Another theoretical line of work considers the verification problem for runs starting from a {\em fixed} initial database.
During the run, the database may evolve via updates, insertions and deletions. 
Since inputs may contain fresh values from an infinite domain, this verification variant remains infinite-state.
The property languages are fragments of first-order-extended $\mu$-calculus~\cite{DBLP:journals/ijcis/GiacomoMR12}.
Decidability results are based on sufficient syntactic restrictions ~\cite{DBLP:journals/ijcis/GiacomoMR12,DBLP:conf/pods/HaririCGDM13,calvanese2015verification}. \cite{DBLP:conf/icsoc/BelardinelliLP12} derives decidability of the verification variant
by also disallowing unbounded accumulation of input values, but this condition is postulated as a semantic property (shown undecidable in \cite{DBLP:conf/pods/HaririCGDM13}). \cite{abdulla2016recency} takes a different approach, in which
decidability is obtained for {\em recency-bounded} artifacts, in which only recently introduced values are retained in the current 
data.

On the practical side of artifact verification, \cite{rawsys} considers the verification of business processes specified in 
a Petri-net-based model extended with data and process components, 
in the spirit of the theoretical work of \cite{rosa2011decidability, badouel2015petri, lazic2008nets, sidorova2011soundness},
which considers extending Petri nets with data-carrying tokens.
The verifier of \cite{rawsys} differs fundamentally from ours in that properties are checked only for a given initial database. In contrast, our verifier checks that {\em all} runs 
satisfy given properties {\em regardless} of the initial underlying database.
\cite{gsmc} and its prior work \cite{gonzalez2012verifying,gonzalez2013model} implemented 
a verifier for artifact systems specified directly in the GSM model.
While the above models are expressive, the verifiers require
restrictions of the models strongly limiting modeling power \cite{gonzalez2012verifying}, 
or predicate abstraction resulting in loss of soundness and/or completeness \cite{gonzalez2013model,gsmc}. Lastly, the properties verified in \cite{gonzalez2013model,gsmc} focus on temporal-epistemic properties in a multi-agent finite-state system. Thus, the verifiers in these works have a different focus and are incomparable to ours.

Practical verification has also been studied in business process management
(see \cite{bpm-verification} for a survey). The considered models are mostly process-driven 
(BPMN, Workflow-Net, UML etc.), with the business-relevant data abstracted away.
The implementation of a verifier for data-driven web applications was studied in \cite{wave2005, wave2006}.
The model is similar in flavor to the artifact model, but much less expressive. The verification approach
developed there is not applicable to HAS*, which requires substantially new tools and techniques.
Finally, our own work on building a verifier based on Spin was discussed in Section \ref{sec:intro},
and compared to VERIFAS in Section \ref{sec:experiment}.

%% file: conclusion.tex
\section{Conclusion} \label{sec:conclusion}

We presented the implementation of VERIFAS, an efficient verifier of temporal properties for 
data-driven workflows specified in HAS*, a variant of the Hierarchical Artifact System model
studied theoretically in~\cite{pods16}. HAS* is inspired by the Business Artifacts framework introduced 
by IBM~\cite{GSM:DEBS-2011} and incorporated in OMG's CMMN standard~\cite{GSM-CMMN:2012,OMG:CMMN:Beta1}.

While the verification problem is {\sc expspace}-complete (see extended version \cite{extended}) 
our experiments show that the theoretical worst case is unlikely in practice
and that verification is eminently feasible. Indeed, VERIFAS achieves excellent performance 
(verification within seconds) on a practically relevant class of real-world and synthetic workflows 
(those with cyclomatic complexity in the range recommended by good software engineering practice),
and a set of representative properties.
The good performance of VERIFAS is due to an adaptation of our symbolic verification techniques
developed in \cite{pods16}, coupled with the classic Karp-Miller algorithm
accelerated with an array of nontrivial novel optimizations.

We also compared VERIFAS to a verifier we built on top of the widely used model checking tool Spin.
VERIFAS not only applies to a much broader class of artifacts but also outperforms the Spin-based verifier
by over one order of magnitude
even on the simple artifacts the Spin-based verifier is able to handle.
To the best of our knowledge, VERIFAS is the first implementation of practical significance of an artifact verifier 
with full support for unbounded data. In future work, we plan to extend VERIFAS to support
a more expressive model that captures true parallelism, aggregate functions and arithmetic.

%% file: appendix-semantics.tex
\section{Some Formal Definitions} \label{app:semantics}

In this section we provide further details on the syntax and semantics of HAS*, the syntax of LTL-FO, and symbolic transitions.

\vspace{4mm}
\noindent
{\bf Opening and closing services} \\
We begin with the formal definition of opening and closing services of tasks, described informally in the paper.

\victor{Changed the definition to refer explicitly to the input and output variables that are now part of task schemas. Also, the direction of the $f_{out}$ mapping was changed (seems more intuitive).}
\begin{definition}
Let $T_c$ be a child of a task $T$ in $\cala$. \\
(i) The \textbf{opening-service} $\sigma_{T_c}^o$ of $T_c$ is a tuple $\langle 
\pi, f_{in} \rangle$, where $\pi$ is a condition over $\bar{x}^{T}$, and
$f_{in}$ is a 1-1 mapping from $\bar{x}^{T_c}_{in}$ to $\bar{x}^{T}$ (called the input variable mapping).
We denote $range(f_{in})$ by  $\bar x^{T}_{{T_c}^\downarrow}$ (the variables of $T$ passed as input to $T_c$).
\\
(ii) The \textbf{closing-service} $\sigma_{T_c}^c$ of $T_c$ is a tuple $\langle 
\pi, f_{out} \rangle$,
where $\pi$ is a condition over $\bar{x}^{T_c}$,
and $f_{out}$ is a 1-1 mapping from $\bar{x}^{T_c}_{out}$ to $\bar{x}^{T}$ (called the output variable mapping).
We denote $range(f_{out})$ by $\bar{x}^{T}_{{T_c}^\uparrow}$, referred to as the \textbf{returned variables} from $T_c$.
It is required that
$\bar{x}_{{T_c}^\uparrow}^{T} \cap {\bar x}_{in}^{T} = \emptyset$.
\end{definition}
\vspace{-2mm}


\vspace{4mm}
\noindent
{\bf Semantics of HAS*} \\
We next define the semantics of HAS*. Intuitively, a run of a HAS* on a database $D$ consists of 
an infinite sequence of transitions among HAS* instances (also referred to as configurations, or snapshots), 
starting from an initial artifact tuple satisfying pre-condition $\Pi$, and empty artifact relations. 
We begin by defining single transitions. The intuition is that at each snapshot,
each task $T$ can open a subtask $T_c$ if the pre-condition of the opening service of $T_c$ holds, and the values of
a subset of $\bar{x}^T$ is passed to $T_c$ as its input variables. $T_c$ can be closed if the pre-condition of
its closing service is satisfied. When $T_c$ is closed, the values of a subset of $\bar{x}^{T_c}$ is sent to $T$ as
$T$'s returned variables from $T_c$.
An internal service of $T$ can only be applied after all active sub-tasks of $T$ have returned their answer.

For a mapping $M$, we denote by $M[a \mapsto b]$ the mapping that sends $a$ to $b$ and agrees with $M$ everywhere else.
\begin{definition}
Let $\Gamma = \langle \mathcal{A}, \Sigma, \Pi \rangle$ be a hierarchical artifact system, where
$\mathcal{A} = \langle \calh, \mathcal{DB} \rangle$. We define the transition relation
$(\nu, stg, D, S) \goto{\sigma} (\nu', stg', D', S')$
among instances of $\cala$, where $\sigma$ is a service, as follows.
For an internal service $\sigma = \langle \pi, \psi, \bar{y}, \delta \rangle$ of task
$T$, $(\nu, stg, D, S) \goto{\sigma} (\nu', stg', D', S')$ if
$D = D'$ and the following hold:
\begin{itemize}\itemsep=0pt\parskip=0pt
\item $stg(T) = \aactive$, $D \models \pi(\nu)$ and $D \models \psi(\nu')$,
\item $stg(T_c) = \aclosed$ for every $T_c \in child(T)$,
\item $\nu'(y) = \nu(y)$ for each $y \in \bar{y}$,
\item $\nu'(y) = \nu(y)$ for each $y \in \bar{x}^{\tilde{T}}$ for $\tilde{T} \neq T$,
\item $stg' = stg$, 
\item if $\delta = \{+\cals_i(\bar{z})\}$, then $S' = S[\cals_i \mapsto S(\cals_i) \cup \{\nu(\bar{z})\}]$,
\item if $\delta = \{-\cals_i(\bar{z})\}$, then
$\nu'(\bar{z}) \in S(\cals_i)$ and $S' = S[\cals_i \mapsto S(\cals_i) - \{\nu'(\bar{z})\}]$,
\item if $\delta = \emptyset$ then $S' = S$.
\end{itemize}

If $\sigma =  \sigma^o_T = \langle \pi, f_{in} \rangle$ is the opening-service for a non-root task $T$,
then $(\nu, stg, D, S) \goto{\sigma} (\nu', stg', D', S')$ if
$D = D'$ and:
\begin{itemize}\itemsep=0pt\parskip=0pt
\item $stg(T) = \aclosed$, $stg(T_p) = \aactive$ where $T_p$ is the parent of $T$,
and $D \models \pi(\nu)$
\item $\nu' = \nu[y \mapsto \nu(f_{in}(y)), z \mapsto \anull, y \in \bar{x}^T_{in}, z \in 
(\bar{x}^T - \bar{x}^T_{in})]$
\item $stg' = stg[T \mapsto \aactive]$, and
\item $S' = S[\cals^T \mapsto \emptyset]$.
\end{itemize}

If $\sigma = \sigma^c_T = \langle \pi, f_{out} \rangle$ is the closing-service for a non-root task $T$,
then $(\nu, stg, D, S) \goto{\sigma} (\nu', stg', D', S')$ if $D = D'$ and:
\begin{itemize}\itemsep=0pt\parskip=0pt
\item $stg(T) = \aactive$ and $D \models \pi(\nu)$,
\item $stg(T_c)  = \aclosed$ for every $T_c \in child(T)$,
\item $\nu' = \nu[f_{out}(y) \mapsto \nu(y)\mid y \in \bar{x}^{T}_{out}]$,  \victor{Modified to fit the new $f_{out}$ definition.}
\item $stg' = stg[T \mapsto \aclosed]$, and $S' = S[\cals^T \mapsto \emptyset]$. 
\end{itemize}

\end{definition}

We next define {\em runs} of artifact systems. We will assume that runs are {\em fair},
i.e. no task is starved forever by other running tasks. Fairness is commonly ensured by schedulers in multi-process systems.
We also assume that runs are non-blocking, i.e. for each task that has not yet returned its answer,
there is a service applicable to it or to one of its descendants.
For an artifact schema $\mathcal{A} = \langle \calh, \db \rangle$ and a node $T$ of $\calh$, 
we denote by $\emph{tree(T)}$ the subtree of $\calh$ rooted at $T$, 
$\emph{child}(T)$ the set of children of $T$ (also called {\em subtasks} of $T$), 
$\emph{desc}(T)$ the set of descendants of $T$ (excluding $T$). Finally, $\emph{desc}^*(T)$ denotes
$\emph{desc}(T) \cup \{T\}$.

\begin{definition}
Let $\Gamma = \langle \mathcal{A}, \Sigma, \Pi \rangle$ be an artifact system,
where $\mathcal{A} = \langle \calh, \mathcal{DB} \rangle$.
A \emph{run} of $\Gamma$ on database instance $D$ over $\mathcal{DB}$
is an infinite sequence $\rho = \{ (I_i, \sigma_i) \}_{i \geq 0}$,
where each $I_i$ is an instance $(\nu_i, stg_i, D, S_i)$ of $\cala$, $\sigma_i \in \Sigma$,
$\sigma_0 = \sigma^o_{T_1}$, $D \models \Pi(\nu_0)$, $stg_0 = \{T_1 \mapsto \aactive, T_i \mapsto \aclosed \mid 2 \leq i \leq k\}$, 
$S_0 = \{\cals_{\calh} \mapsto \emptyset \}$, and for each $i > 0$
$I_{i-1} \goto{\sigma_{i}} I_{i}$. In addition,
for each $i \geq 0$ and task $T$ active in $I_i$, there exists $j > i$ such that
$\sigma_j \in \bigcup_{T' \in \emph{desc}^*(T)} \Sigma^{oc}_{T'}$.
\end{definition}

We denote by $Runs(\Gamma)$ the set of runs of $\Gamma$.
Observe that all runs of $\Gamma$  are infinite.
In a given run, the root task itself may have an infinite run,
or other tasks may have infinite runs.
However, if a task $T$ has an infinite run,
then none of its ancestor tasks can make an internal transition or return
(although they can still call other children tasks).

Because of the hierarchical structure of HAS*, and the locality of task specifications,
the actions of independent tasks running concurrently can be arbitrarily interleaved.
In order to express properties of HAS* in an intuitive manner, it will be useful to ignore such interleavings and
focus on the {\em local runs} of each task, consisting of the transitions affecting the local variables and artifact 
relations of the task, as well as interactions with its children tasks. More precisely, let 
$\Gamma = \langle \mathcal{A}, \Sigma, \Pi \rangle$ be an artifact system, 
$T =  \langle \bar{x}^T, \cals^T \rangle$ a task in $\Gamma$, and $\rho$ a run of $\Gamma$.  
A local run of $T$ induced by $\rho$ is a sequence $\rho_T = \{(I_i,\sigma_i)\}_{i < \gamma}$, where:

\begin{itemize}\itemsep=0pt\parskip=0pt
\item $\gamma \in \mathbb{N} \cup \{\omega\}$
\item $\{\sigma_i\}_{i < \gamma}$ is a sequence of consecutive services in $\Sigma^{obs}_T$ occurring in $\rho$, 
of which $\sigma_0 = \sigma^o_T$ 
\item if $\gamma \in  \mathbb{N}$, then $\sigma_{\gamma - 1} = \sigma_T^c$ and $\sigma_i \neq \sigma_T^c$ for $0 < i < \gamma -1$
\item if $\gamma = \omega$, then $\sigma_i \neq \sigma_T^c$ for $0 < i < \omega$
\item $\{I_i\}_{i < \gamma}$ are the consecutive instances of $T$ induced by the services  $\{\sigma_i\}_{i < \gamma}$ in the run $\rho$
\end{itemize}
We denote by $Runs_T(\rho)$ the set of local runs of $T$ induced by the run $\rho$ of $\Gamma$, and
$Runs_T(\Gamma) = \bigcup_{\rho \in Runs(\Gamma)}~ Runs_T(\rho)$.

\vspace{4mm}
\noindent
{\bf LTL-FO syntax} \\
We next provide the definition of LTL-FO formulas.
\begin{definition}
Let $T$ be a task of a HAS* $\Gamma$.
Let $\bar y$ be a finite sequence of variables in $\varid \cup \varnum$ disjoint from $\bar{x}^T$, called {\em global variables}.
An LTL-FO formula for $T$ is an expression $\forall \bar y \varphi_f$, where:
\vspace*{-1mm}
\begin{itemize}\itemsep=0pt\parskip=0pt
\item $\varphi$ is an LTL formula with propositions $P \cup \Sigma_T^{obs}$,
where $P$ is a finite set of propositions \yuliang{fixed typo} disjoint from $\Sigma_T^{obs}$
\item $f$ is a function from $P$ to conditions over $\bar x^T \cup \bar y$
\item $\varphi_f$ is obtained by replacing each $p \in P$ with $f(p)$
\end{itemize}
\end{definition}

\vspace{4mm}
\noindent
{\bf Symbolic Transitions}  \\
Finally, we provide the formal definition of symbolic transitions.
Although our discussion in Section \ref{sec:karp-miller} focuses primarily on single tasks, 
the notion of partial symbolic instances and symbolic transitions 
can be naturally extended to the full HAS* model. 
For clarity, we first define symbolic transitions for single tasks (ignoring interactions with children) 
and then extend the definition to multiple tasks (taking into account such interactions). 

\vspace*{1mm}
\noindent
\textit{Single Tasks}

\noindent
For a single task, we define $\asucc(\tau, \bar{c})$ as follows. 
For a given quantifier-free FO formula $\varphi$, we denote by $\aflat(\varphi)$
the formula obtained by replacing each relational atom $R(x, y_1, \dots, y_m)$ with 
$\bigwedge_{i=1}^m x.A_i = y_i$, and denote by $\aconj(\varphi)$ the set of conjuncts of the 
disjunctive normal forms of $\aflat(\varphi)$. Note that each literal in every $\aconj(\varphi)$
is positive as negations can be removed by inverting each $=$ and $\neq$.
We let $t(\theta)$ to be the partial isomorphism type
induced by $\theta$ for $\theta \in \aconj(\varphi)$. 
Given a PSI $I = (\tau, \bar{c})$,
for every service $\sigma = (\pi, \psi, \bar{y}, \delta)$, for every $\theta_1 \in \aconj(\pi)$ where
$\tau \cap t(\theta_1)$ is satisfiable (the pre-condition is satisfied), 
for every $\theta_2 \in \aconj(\psi)$, $\asucc(I)$ contains
the following PSIs that are valid:
\begin{itemize}\itemsep=0pt\parskip=0pt
\item $((\tau \cap t(\theta_1) | \bar{y}) \cap t(\theta_2), \bar{c})$ if $\delta = \emptyset$,
\item $((\tau \cap t(\theta_1) | \bar{y}) \cap t(\theta_2), 
\bar{c}[\tau_S \rightarrow \bar{c}(\tau_S) + 1])$ for $\tau_S = f_{\bar{z} \rightarrow \cals}(\tau | \bar{z})$ if $\delta = \{+S(\bar{z})\}$, or
\item $((\tau \cap t(\theta_1) | \bar{y}) \cap t(\theta_2) \cap f_{\cals \rightarrow \bar{z}}(\tau_S), 
\bar{c}[\tau_S \rightarrow \bar{c}(\tau_S) - 1])$ if $\delta = \{-S(\bar{z})\})$ for some $\tau_S$ where $\bar{c}(\tau_S) > 0$.
\end{itemize}

\vspace*{1mm}
\noindent
\textit{Extension to Multiple Tasks}

\noindent
For each task $T$, the definition of partial isomorphism type of $T$ is the same as 
Definition \ref{def:partial-isomorphism-type}, except the set $\cale$ is replaced with
$\cale^T$, the set of all expressions of task $T$. Partial symbolic instances are defined as follows.

\begin{definition} \label{def:psi-full}
A \textit{partial symbolic instance} (PSI) $I$ of a task $T$ is a tuple 
$(\tau, \bar{c}, \bar{r})$ where 
\begin{itemize}\itemsep=0pt\parskip=0pt
\item $\tau$ is a partial isomorphism type of $T$,
\item $\bar{c}$ is a vector of $\mathbb{N}$ where 
each dimension of $\bar{c}$ corresponds to a unique partial isomorphism type of $T$, and
\item $\bar{r}$ is a mapping from $child(T)$ to the set $\{\mathtt{active}, \mathtt{inactive}\}$. 
\end{itemize}
\end{definition}

Intuitively, the extra component $\bar{r}$ records the status and return information of
child tasks of $T$. For a child task $T_c \in child(T)$, $\bar{r}(T_c)$ indicates whether
$T_c$ is active or not. 

Symbolic transitions are specified by the successor function  \\
$\asucc^T(I)$ for each task $T$,
which is formally defined as follows. We first consider successors under internal services, then under
opening and closing services of the children of $T$.

\victor{If we take the simpler definition that $\bar{r}$ only needs to provide the status of each child task,
active or inactive. I assume this below (but did not change it above).}
\yuliang{Okay. I changed the above to match this.}

\vspace*{1mm}
\noindent
\textit{Internal Services} Given a PSI $I = (\tau, \bar{c}, \bar{r})$ of task $T$
the definition of symbolic transitions with internal services resembles the definition 
in the single task case, with the extra conditions that 
$\bar{r}(T_c)  = \aclosed$ for every child task $T_c \in child(T)$ and for every resulting 
PSI $(\tau', \bar{c}', \bar{r}')$,
$\bar{r}' = \bar{r}$.

\vspace*{1mm}

\victor{Here is an attempt at the simpler definition, replacing what is below( up to remarks on monotone prunning).}
\yuliang{Looks good. Added the check for the opening condition}

We next consider opening and closing services of $T$'s children.
For PSI $I = (\tau, \bar{c}, \bar{r})$ of task $T$,
$\asucc^T(I)$ contains the following:

\vspace*{1mm}
\noindent
\textit{Opening service $\sigma^o_{T_c}$ with pre-condition $\pi$: } 
(applicable if $\bar{r}(T_c) = \aclosed$)
$$\asucc^T(I) \mbox{ contains } (\tau \cap t(\theta), \bar{c}, \bar{r}[T_c \mapsto \aactive])$$
for every $\theta \in \aconj(\pi)$ such that $\tau \cap \theta$ is satisfiable.

\vspace*{2mm}
\noindent
\textit{Closing service $\sigma^c_{T_c}$: } (applicable if $\bar{r}(T_c) = \aactive$)
$$\asucc^T(I) \mbox{ contains } 
(\tau \cap \tau_c, \bar{c}, \bar{r}[T_c \mapsto \aclosed])$$
for every partial isomorphism type $\tau_c$ of $\bar{x}^{T}_{{T_c}^\uparrow}$ such that
$\tau \cap \tau_c$ is satisfiable.
\vspace*{1mm}

%% file: appendix-example.tex
\section{Running Example} \label{app:example}

We provide our running example of a HAS* specification for an order fulfillment business process
based on a real-world BPMN workflow in \cite{bpmn} (the ``Order Fulfillment and Procurement'' workflow under tab ``Example'').
The workflow allows the customer to place orders
and the supplier company to process the orders. It has the following database schema:

\vspace{-1mm}
\begin{itemize}\itemsep=0pt\parskip=0pt
\item
\dbcustomers $\mathtt{(\underline{\emph{ID}}, name, address, record)}$\\
\dbitems $\mathtt{(\underline{\emph{ID}}, item\_name, price)}$\\
\dbrecords $\mathtt{(\underline{\emph{ID}}, status)}$\\
\end{itemize}
\vspace{-5mm}

In the schema, the IDs are key attributes, 
$\mathtt{price}$, $\mathtt{item\_name}$, $\mathtt{name}$, $\mathtt{address}$, $\mathtt{status}$
are non-key attributes, and $\mathtt{record}$ is a foreign key attribute
satisfying the dependency \\ $\dbcustomers[record] \subseteq \dbrecords[\emph{ID}]$.

Intuitively, the $\dbcustomers$ table contains customer information with a foreign key pointing to
the customers' credit records stored in $\dbrecords$. 
The $\dbitems$ table contains information of all the items. Note that the schema is acyclic as there is only
one foreign key reference from $\dbcustomers$ to $\dbrecords$.

\begin{figure}[!ht]
\vspace{-2mm}
\centering
\includegraphics[scale=0.5]{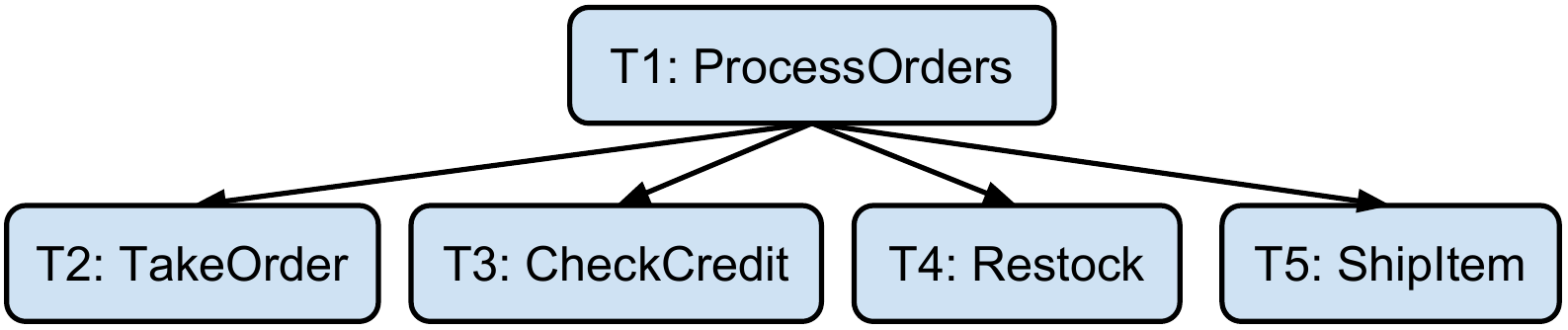}
\vspace{-2mm}
\caption{Tasks Hierarchy} 
\label{fig:hierarchy2}
\vspace{-2mm}
\end{figure}

The artifact system has 5 tasks: $T_1$:\textbf{ProcessOrders}, \\
$T_2$:\textbf{TakeOrder}, $T_3$:\textbf{CheckCredit}, $T_4$:\textbf{Restock} and \\ $T_5$:\textbf{ShipItem},
which form the hierarchy represented in Figure \ref{fig:hierarchy2}.
Intuitively, the root task \textbf{ProcessOrders} serves as a global coordinator which
maintains a pool of all orders and the child tasks \textbf{TakeOrder}, 
\textbf{CheckCredit}, \textbf{Restock} and \textbf{ShipItem} serve as the 4 sequential stages of an order. 
At a high level, \textbf{ProcessOrders} repeatedly picks an order from its pool and
processes it with a stage by calling the corresponding child task. 
After the child task returns, the order is either placed back to the pool or processed with the next stage.
For each order, the workflow first takes the customer and item information with the \textbf{TakeOrder} task.
The credit record of the customer is checked by the \textbf{CheckCredit} task. If the record is good, 
then \textbf{ShipItem} can be called to ship the item to the customer. If the requested item is unavailable,
then \textbf{Restock} must be called before \textbf{ShipItem} to procure the item.

Next, we describe each task in detail. For convenience, 
we use existential quantification in conditions, which can be simulated by adding extra variables.
Also for convenience, when we specify that a variable of a child task is an input/output variable, 
it is mapped to the variable of the same name in the parent task. We also omit specifying the type of
the variable because by definition it has the same types as the variable that it maps to.


\vspace{1.5mm}
\noindent
\textbf{ProcessOrders}: 
As the root task, its opening condition is $\mathtt{True}$ and closing condition is $\mathtt{False}$.
The task has the following artifact variables (none of which are input or output variables):

\vspace{-1mm}
\begin{itemize}\itemsep=0pt\parskip=0pt
\item ID variables: $\acustomerid$ of type $\dbcustomers.\emph{ID}$ and $\aitemid$ of type $\dbitems.\emph{ID}$
\item Non-ID variables: $\astatus$ and $\ainstock$
\end{itemize}
\vspace{-1mm}

All variables are initialized to $\anull$ by the global pre-condition.
The task also has an artifact relation \\ $\mathtt{ORDERS}(\acustomerid, \aitemid, \astatus, \ainstock)$ 
with attributes of the same types as the variables.
Intuitively, $\mathtt{ORDERS}$ stores the orders to be processed, where each order consists of
a customer and an ordered item. The variable $\astatus$ indicates the current status of the order
and $\ainstock$ indicates whether the item is currently in stock.

The task has 3 internal services: \emph{Initialize}, \emph{StoreOrder} and \emph{RetrieveOrder}.
Intuitively, \emph{Initialize} creates a new order with \\ $\acustomerid = \aitemid = \anull$.
When \emph{RetrieveOrder} is called, an order is non-deterministically chosen and removed
from $\mathtt{ORDERS}$ for processing, and $(\acustomerid, \aitemid, \astatus, \ainstock)$ is set to be the chosen tuple.
When \emph{StoreOrder} is called, the current order \\
$(\acustomerid, \aitemid, \astatus, \ainstock)$ is inserted into
$\mathtt{ORDERS}$. The latter two services are specified as follows.

\vspace{2mm}
\noindent
\emph{StoreOrder}: \\
Pre: $\acustomerid \neq \anull \land \aitemid \neq \anull \land \astatus \neq \text{``Failed"}$ \\
Post: $\acustomerid = \anull \land \aitemid = \anull \land \astatus = \text{``Init"}$ \\
Update: $\{+\mathtt{ORDERS}(\acustomerid, \aitemid, \astatus, \ainstock)\}$

\vspace{2mm}
\noindent
\emph{RetrieveOrder}: \\
Pre: $\acustomerid = \anull \land \aitemid = \anull$ \\
Post: $\mathtt{True}$ \\
Update: $\{-\mathtt{ORDERS}(\acustomerid, \aitemid, \astatus, \ainstock)\}$

\vspace{1mm}
The sets of propagated variables are empty for both services,
 since \textbf{ProcessOrders} has no input variable.  

\vspace{2mm}
\noindent
\textbf{TakeOrder}: When this task is called, the customer enters 
the information of the order ($\acustomerid$ and $\aitemid$) and 
the status of the order is initialized to $\text{``OrderPlaced"}$.
The task contains $\acustomerid$, $\aitemid$, $\astatus$ and $\ainstock$ as variables
and all are output variables to the parent task. 
There are two internal services,
\emph{EnterCustomer} and \emph{EnterItem}, which allow the customer to enter 
his/her information and the item's information. The $\dbcustomers$ and $\dbitems$
tables are queried to obtain the customer ID and item ID. 
When \emph{EnterItem} is called, the supplier checks whether the item is currently in stock and
sets the variable $\ainstock$ to ``Yes'' or ``No'' accordingly.
These two services can be called multiple times to allow the customer to modify previously 
entered data. 
The closing condition of this task is $\acustomerid \neq \anull \land \aitemid \neq \anull$.
Formally, the \emph{EnterCustomer} service is specified as follows and 
the \emph{EnterItem} service is similar: 

\vspace{1mm}
\noindent
\emph{EnterCustomer}:  \\
Pre: $\mathtt{True}$ \\ 
Post:
\vspace*{-1.5mm}
\begingroup
\addtolength{\jot}{-0.7mm}
\begin{align*}
& \exists n \exists a \exists r \ \dbcustomers(\acustomerid, n, a, r) \land ((\acustomerid \neq \anull \land \\
& \aitemid \neq \anull) \rightarrow \astatus = \text{``OrderPlaced"} ) \land \\
& ((\acustomerid = \anull \lor \aitemid = \anull) \rightarrow \astatus = \anull ) 
\end{align*}
\endgroup

\vspace*{-1mm}
\noindent
Propagate: $\{\ainstock, \aitemid\}$

\vspace*{2mm}

\noindent
\textbf{CheckCredit}: This task checks the financial record of a customer and 
decides whether the supplier should sell to the customer.
It has opening condition $\astatus = \text{``OrderPlaced"}$.
When the credit record is good, $\astatus$ is set to $\text{``Passed"}$
and otherwise set to $\text{``Failed"}$. After $\astatus$ is set,
the task can close and output $\astatus$ to the parent task.
This task contains the following variables:
\vspace{-1mm}
\begin{itemize}\itemsep=0pt\parskip=0pt
\item ID variables:  $\acustomerid$ (input variable from $T_1$), $\arecord$ of type $\dbrecords.\emph{ID}$
\item Non-ID variables: $\astatus$ (output variable).
\end{itemize}
\vspace{-1mm}
The task has a single internal service \emph{Check} to perform the credit check, specified as follows.

\vspace{1mm}
\noindent
\emph{Check}: \\
Pre: $\mathtt{True}$ \\
Post:
\vspace*{-1mm}
\begingroup
\addtolength{\jot}{-0.7mm}
\begin{align*}
& \exists n \exists a \ \dbcustomers(\acustomerid, n, a, \arecord) \land \\
& (\dbrecords(\arecord, \text{``Good"}) \rightarrow \astatus = \text{``Passed"}) \land \\
& (\neg \dbrecords(\arecord, \text{``Good"}) \rightarrow \astatus = \text{``Failed"}) 
\end{align*}
\endgroup

\vspace*{-1mm}
\noindent
Propagate: $\{\acustomerid\}$

\vspace*{2mm}
\noindent
\textbf{Restock} and \textbf{ShipItem}: 
The \textbf{Restock} task can open if the requested item is out of stock ($\ainstock = \text{``No''}$).
It takes $\aitemid$ as input and returns $\ainstock$ as output.
The task has a single internal service \emph{Procure} that can be called repeatedly 
until the procurement is successful ($\ainstock$ is set to ``Yes" and returned to $T_1$).

The \textbf{ShipItem} task can open only when the credit check is passed ($\astatus = \text{``Passed"}$)
and the item is available ($\ainstock = \text{``Yes"})$. 
If the shipment is successful then $\astatus$ is set to $\text{``Shipped"}$ 
otherwise the order fails ($\astatus$ is set to $\text{``Failed"}$). 
These two tasks are specified similarly to \textbf{CheckCredit}, so we omit the details.

%% file: appendix-karp-miller.tex
\section{Repeated-Reachability} \label{app:karp-miller}


We describe in more detail how our Karp-Miller based algorithm with pruning can be extended to 
state repeated reachability, thus providing full support for verifying LTL-FO properties.
For VASS, it is well-known that the coverability set $\cali$ extracted from the tree $\calt$
constructed by the classic Karp-Miller algorithm can be used to 
identify the repeatedly reachable partial isomorphism types (see Section~\ref{sec:classic-km}).
The same idea can be extended to the Karp-Miller algorithm with monotone pruning (Section \ref{sec:pruning}),
since the algorithm guarantees to construct a coverability set. This can be done as follows.

First, for every $I = (\tau, \bar{c}) \in \cali$,
if there exists $\tau_S$ such that $\bar{c}(\tau_S) = \omega$,
then $I$ is inherently repeatedly reachable because 
the accelerate operation was applied to generate the $\omega$.
In addition, it is sufficient to consider only the
\emph{maximal} PSIs in $\cali$. A PSI $I$ is maximal if there is no
state $I' \neq I$ in $\cali$ and $I \preceq I'$. 
We denote by $\imax$ the subset of maximal PSIs in $\cali$ that contains no $\omega$.

Then for PSI $I \in \imax$, it is not difficult to show that 
$I$ is repeatedly reachable iff $I$ is contained in a cycle consisting of \emph{only} PSIs in $\imax$.
As a result, the set of repeatedly reachable states can be computed by
constructing the transition graph among $\imax$ and computing its \emph{strongly connected components}.
A PSI $I$ is repeatedly reachable if $I$ is contained in a component with at least 1 edge.

The reason this works is the following. 
Suppose $I \in \imax$ and $I$ is contained in a cycle $\{I_i\}_{a \leq i \leq b}$ where $I_a$ is not maximal. 
Then there exists a reachable PSI $I_a'$ where $I_a < I_a'$.
So the sequence of transitions that leads $I_a$ to $I_b$ is also a valid sequence 
that can be applied starting from $I_a'$ and results in a cycle $\{I_i'\}_{a \leq i \leq b}$.
The transitions leave the counters unchanged, so $I_i' > I_i$ for every $i \in [a, b]$,
which contradicts the assumption that $I$ is maximal.

However, the same approach cannot be directly applied when the Karp-Miller algorithm 
is equipped with our $\preceq$-based pruning (Section~\ref{sec:aggressive-pruning}).
The algorithm explores fewer states due to the more aggressive pruning, and
it turns out that the resulting coverability set $\cali^\preceq$ is 
incomplete to determine whether a state is repeatedly reachable.
We can no longer guarantee that a repeatedly reachable state is only contained 
in a cycle of maximal states in $\cali^\preceq$.
The above reasoning fails because it relies on the \emph{strict monotonicity} property:
for every PSI $I$ and $I'$ where $I < I'$, for every $I_{\mathtt{next}} \in \asucc(I)$,
there exists $I_{\mathtt{next}}' \in \asucc(I')$ such that $I_{\mathtt{next}} < I_{\mathtt{next}}'$.
This no longer holds when $<$ is replaced with $\prec$.
Consequently, we need to explore PSIs outside of $\imax$ when we detect cycles.
To avoid state explosion when the extra PSIs are explored, 
we make use of a pruning criterion obtained by slightly restricting $\preceq$ as follows.


\vspace*{-1.5mm}
\begin{definition} \label{def:prec}
Given two partial symbolic states $I = (\tau, \bar{c})$ and $I' = (\tau', \bar{c}')$,
$I \preceq^+ I'$ iff $I = I'$ or the following hold:
\begin{itemize}
\item $\tau \models \tau'$, 
\item there exists $f$ that satisfies the conditions of Definition \ref{def:preceq}, and 
\item there exists $\tau_S'$ such that $\sum_{\tau_S} f(\tau_S, \tau_S') < \bar{c}'(\tau_S')$.
\end{itemize}
\end{definition}
\vspace*{-1.5mm}

By using the same Karp-Miller-based algorithm with $\preceq^+$ in monotone pruning, 
we compute a set $\imax^+$ that satisfies the following:
for every $I \in \imax$ and $I'$ reachable from $I$, there exists $I'' \in \imax^+$ 
such that $I' \preceq^+ I''$. Since $\preceq^+$ satisfies the strict monotonicity property,
a PSI $I \in \imax$ is repeatedly reachable iff $I$ is contained in a cycle that 
contains only PSIs in $\imax^+$, which can be checked efficiently by computing the strongly connected components.
Note that when we compute $\imax^+$, we can accelerate the search by pruning a new PSI $I$
if $I \preceq^+ I'$ for any $I'$ in the previously computed Karp-Miller tree.
In addition, there is no need to apply the accelerate operator since 
it is sufficient to consider only PSIs with no $\omega$. 
As confirmed by our experiments, 
the overhead for computing the set of repeatedly reachable PSIs is acceptable.

%% file: appendix-experiment.tex
\section{Synthetic Workflow Generator} \label{app:experiment}

We briefly describe how the synthetic workflows used in our experiments were produced.
Each part of a synthetic workflow is generated at random,
for the given parameters \#relations, \#tasks, \#variables and \#services.

We first generate a random tree of fixed size as the acyclic database schema 
where each relation has a fixed number of (4) non-ID attributes.
Then we generate a random tree of fixed size as the task hierarchy.
Within each task, for each variable type (non-ID or ID of some relation),
we uniformly generate the same number of variables.
We randomly choose 1/10 of the variables as input variables
and another 1/10 as output variables. 
Then we generate a fixed number of internal services for each task
with randomly generated pre-and-post conditions (described next). With probability 1/3,
the internal service has $\bar{y}$ propagating a randomly chosen subset (1/10) 
of the task's variables, or inserting a fixed tuple of variables into the artifact relation,
or retrieving a tuple from the artifact relation. 

Each condition of each service is also generated as a random tree.
We first generate a fixed number of (5) atoms,
where each atom has 1/3 of probability of having the form
$x = y$, $x = c$ or $R(\bar{x})$, where $x, y, \bar{x}$ are 
variables chosen uniformly at random and $c$ is a random constant from a fixed set.
Each leaf is negated with probability 1/2.
Then we generate the condition as a random binary tree with the atoms as the leafs of the tree.
Each internal node of the binary tree is an $\land$-connective with probability 4/5
and an $\lor$-connective with probability 1/5. 
We chose to generate $\land$ with higher probability 
based on our observations of the real workflows.

%% file: appendix-table.tex
\begin{table*}[ht]
\begin{center}
{\large Table of Symbols} \\
\vspace{4mm}
\begin{tabular}{||ll||}
\hline
\hline
   
$\db$        &    database schema of a HAS*    \\ \hline

$\bar x^T$    & artifact variables of task $T$                    \\ \hline

$\bar{x}^T_{in}$  & input variables of task $T$                    \\ \hline

$\bar{x}^T_{out}$  & output variables of task $T$                    \\ \hline

$\cals^T$   & artifact relations of task $T$                    \\ \hline

$\calh$    & task hierarchy         \\ \hline

$\emph{child}(T)$ & set of children of task $T$             \\ \hline

$\emph{desc}(T)$ & set of descendants of task $T$ (excluding $T$)             \\ \hline

$\emph{desc}^*(T)$  & set of descendants of task  $T$ (including $T$)                    \\ \hline

$\cals_\calh$     &  artifact relations of all tasks in $\calh$   \\ \hline

{\em stg}             & stage mapping from tasks to $\{\aactive, \aclosed \}$ \\ \hline

$\pi$             & pre-condition of an internal task                   \\ \hline
$\psi$             & post-condition of an internal task                   \\ \hline
$\delta$             & update of an internal task                   \\ \hline
$\bar y$         &  propagated variables of an internal task \\ \hline

$\sigma_{T}^o$   & opening service of task $T$                    \\ \hline
$\sigma_{T}^c$   & closing service of task $T$                    \\ \hline

$\Sigma_T$        & set of internal services of task $T$                    \\ \hline

$\Sigma_T^{oc}$ & $\Sigma_T \cup \{\sigma^o_T, \sigma^c_T\}$ \\ \hline
$\Sigma_T^{\emph{obs}}$ & $ \Sigma_T^{oc} \cup \{\sigma^o_{T_c}, \sigma^c_{T_c} \mid T_c \in \emph{child}(T)\}$ \\ \hline

$\cala$         & artifact schema                    \\ \hline

$\Gamma$   &   HAS* \\ \hline

$\Pi$             &   global pre-condition of a HAS* $\Gamma$  \\ \hline

$\rho = \{ (I_i, \sigma_i) \}_{i \geq 0}~~~~~~$    &  run of a HAS*          \\ \hline

$Runs(\Gamma)$ &   set of runs of a HAS* $\Gamma$  \\ \hline
$Runs_T(\rho)$ &   set of local runs of $T$ induced by the run $\rho$ \\ \hline
$Runs_T(\Gamma)$ &   $\bigcup_{\rho \in Runs(\Gamma)}~ Runs_T(\rho)$ \\ \hline
LTL & linear-time temporal logic \\ \hline
{\bf G},  {\bf F}, {\bf X}, {\bf U} &  ``always", ``eventually", ``next", ``until" \\
 &  LTL temporal operators \\ \hline
$\models$ &  satisfies \\ \hline
$B_\varphi$  &  B\"{u}chi automaton for the LTL formula $\varphi$  \\ \hline
$\omega$  & ordinal omega  \\ \hline
FO  & first-order logic \\ \hline
$\exists$FO & existential FO \\ \hline
LTL-FO &  extension of LTL with quantifier-free FO conditions \\ \hline
VASS  &  vector addition system with states                    \\ \hline
$\cale$  & set of navigation expressions via foreign keys    \\ \hline
$\tau$          & partial isomorphism type  \\ \hline
$(\tau, \bar{c})$   & partial symbolic instance with counters $\bar{c}$  \\ \hline
PSI & partial symbolic instance \\ \hline
$\asucc(I)$ & set of possible successors of a partial symbolic instance $I$             \\ \hline
\hline
\end{tabular}
\end{center}
\end{table*}